# An advanced workflow for single particle imaging with the limited data at an X-ray free-electron laser

Dameli Assalauova[a], Young Yong Kim[a], Sergey Bobkov[b], Ruslan Khubbutdinov[ac], Max Rose[a], Roberto Alvarez[de], Jakob Andreasson[f], Eugeniu Balaur[g], Alice Contreras[hi], Hasan DeMirci[jk], Luca Gelisio[l], Janos Hajdu[fm], Mark S. Hunter[n], Ruslan P. Kurta[o], Haoyuan Li[pn], Matthew McFadden[n], Reza Nazari[dq], Peter Schwander[r], Anton Teslyuk[bs], Peter Walter[n], P. Lourdu Xavier[lnt], Chun Hong Yoon[n], Sahba Zaare[dn], Viacheslav A. Ilyin[bs], Richard A. Kirian[d], Brenda G. Hogue[hiu], Andrew Aquila[n]* and Ivan A. Vartanyants[ac]*

*a Deutsches Elektronen-Synchrotron DESY, Notkestraße 85, Hamburg, D-22607, Germany*

*b National Research Center 'Kurchatov Institute', Akademika Kurchatova pl. 1, Moscow, 123182, Russian Federation*

*c National Research Nuclear University MEPhI (Moscow Engineering Physics Institute), Kashirskoe sh. 31, Moscow, 115409, Russian Federation*

*d Department of Physics, Arizona State University, Tempe, Arizona, AZ 85287, USA*

*e School of Mathematics and Statistical Sciences, Arizona State University, Tempe, Arizona, AZ 85287, USA*

*f Institute of Physics, ELI Beamlines, Academy of Sciences of the Czech Republic, Prague, CZ-18221, Czech Republic*

*g Department of Chemistry and Physics, La Trobe Institute for Molecular Science (LIMS), La Trobe University, Melbourne, Victoria 3086, Australia*

*h School of Life Sciences, Arizona State University, Tempe, Arizona, AZ 85287, USA*

*i Biodesign Institute Center for Immunotherapy, Vaccines and Virotherapy, Arizona State University, Tempe, Arizona, AZ 85287, USA*

*j Stanford PULSE Institute, SLAC National Accelerator Laboratory, 2575 Sand Hill Road, Menlo Park, California, CA 94025, USA*

*k Department of Molecular Biology and Genetics, Koc University, Istanbul, 34450, Turkey*

*l Center for Free Electron Laser Science (CFEL), DESY, Notkestraße 85, Hamburg, D-22607, Germany*

*m Laboratory of Molecular Biophysics, Department of Cell and Molecular Biology, Uppsala University, Husargatan 3, Uppsala, SE-75124, Sweden*

*n SLAC National Accelerator Laboratory, 2575 Sand Hill Road, Menlo Park, California, CA 94025, USA*

*o European XFEL, Holzkoppel 4, Schenefeld, D-22869, Germany*

*p Physics Department, Stanford University, 450 Jane Stanford Way, Stanford, California, CA 94305–2004, USA*

*q School for Engineering of Matter, Transport and Energy, Arizona State University, Tempe, Arizona, AZ 85287, USA*

*r University of Wisconsin Milwaukee, Milwaukee, WI 53211, USA*

*s Moscow Institute of Physics and Technology, Moscow, 141700, Russian Federation*

*t Max-Planck Institute for the Structure and Dynamics of Matter, Luruper Chaussee 149, Hamburg, D-22761, Germany*

*u Biodesign Institute, Center for Applied Structural Discovery, Arizona State University, Tempe, Arizona, AZ 85287, USA*

**Abstract**

An improved analysis for single particle imaging (SPI) experiments, using the limited data, is presented here. Results are based on a study of bacteriophage PR772 performed at the AMO instrument at the Linac Coherent Light Source (LCLS) as part of the SPI initiative. Existing methods were modified to cope with the shortcomings of the experimental data: inaccessibility of information from the half of the detector and small fraction of single hits. General SPI analysis workflow was upgraded with the expectation-maximization based classification of diffraction patterns and mode decomposition on the final virus structure determination step. The presented processing pipeline allowed us to determine the three-dimensional structure of the bacteriophage PR772 without symmetry constraints with a spatial resolution of 6.9 nm. The obtained resolution was limited by the scattering intensity during the experiment and the relatively small number of single hits.

## 1. Introduction

The recent worldwide outbreak of the COVID-19 pandemic has indicated urgency for the development of complementary imaging techniques for the study of virus structures at high resolution. Currently, cryogenic electron microscopy (Cryo-EM) outperforms other methods of analysis of single viruses particles, although to determine the structure one needs to freeze the samples at cryogenic temperatures. Understanding the nature of the same viruses at room temperature as well as in an aerosols may provide information on the actual size distribution of virus particles in aerosols, a factor that may affect infection of humans by these viruses.

As suggested about two decades ago, intense femtosecond pulses from X-ray Free-Electron Lasers (XFELs) may outrun conventional radiation damage, found at synchrotron X-ray sources, or the Coulomb explosion, found at the intensities of XFELs, and therefore allow atomic-resolution structure determination of isolated macromolecules in their native environment at room temperatures (Neutze *et al.*, 2000). After the start of the first hard X-ray FELs (Emma *et al.*, 2010; Ishikawa *et al.*, 2012), it has become clear that atomic resolution can be achieved with intense X-ray pulses from an XFEL on micro- and nano-crystals of proteins, leading to the serial femtosecond crystallography (SFX) technique (Chapman *et al.*, 2011; Aquila *et al.*, 2012; Boutet *et al.*, 2012). However, progress toward high-resolution three-dimensional (3D) electron density images of non-crystalline biological particles using the Single Particle Imaging (SPI) approach has been slow compared to SFX (Bogan et al. 2008; Seibert *et al.*, 2011; Hantke *et al.*, 2014; van der Schot *et al.*, 2015; Ekeberg *et al.*, 2015).

There are several reasons for the lower resolution achieved in SPI experiments in comparison to the SFX technique. The most important include: lack of crystalline periodicity to amplify the signal, weak single particle signal from non-periodic nanoscale objects compared to instrumental background, limited number of usable data frames collected, and heterogeneity of the samples. Radiation damage processes initiated by X-ray photoionization may also play an important role at high resolution, but this limitation is minimized at XFEL sources by performing experiments in the "diffract and destroy" mode (Chapman *et al.*, 2006). One may expect that in order to further enhance resolution, one could increase the power of XFEL pulses to boost the signal-to-noise ratio for bio-particles such as single proteins or virus particles that scatter very weakly. Unfortunately, increased XFEL fluence strips electrons from atoms more efficiently and the scattered power from the bound electrons (that contain structural information) does not increase in proportion to the X-ray fluence (Lorenz *et al.*, 2012; Gorobtsov *et al.*, 2015). Decreasing the pulse length below one femtosecond may help to outrun Auger decay following the photoionization, but further decrease of the pulse duration may not lead to the desired suppression of all ionization channels (Gorobtsov *et al.*, 2015).

Beside these fundamental limitations, there are a few other limiting factors. Background scattering originating from the beamline obscures the weak scattering signals from biological samples, and fluctuations of the beam position and intensity can cause additional variables that need to be accounted for in reconstruction algorithms. Moreover, single particle sample delivery remains a challenging topic. In addition, despite strong development of detector technology (see, for example, (Allahgholi *et al.*, 2019)), the dynamic range of the present detectors is still not sufficient to capture a full diffraction pattern in a strong single shot. To tackle all these limitations of the SPI technique and to push the methodology further, a dedicated SPI consortium was formed at the Linac Coherent Light Source (LCLS) (Aquila *et al.*, 2015).

Several results were reported from the SPI consortium, covering both hard and soft X-ray experiments and focused primarily on well-characterized viruses with sizes of few tens of nanometers. Different analysis methods have been applied to virus structure determination (see, for example, (Munke *et al.*, 2016; Reddy *et al.*, 2017; Kurta *et al.*, 2017; Rose *et al.*, 2018; Shi *et al.*, 2019)). In the present work, we advance the previous studies performed on bacteriophage PR772 (Rose *et al.*, 2018) by extending the general analysis pipeline of the SPI data as first proposed by Gaffney & Chapman, 2007. Main upgraded features include: a classification method based on the expectation-maximization (EM) algorithm, developed in Cryo-EM (Dempster *et al.*, 1977) and mode decomposition for the final virus structure determination (Vartanyants & Singer, 2010; Thibault & Menzel, 2013). EM-based algorithms were first applied in SPI data analysis for 3D orientations recovery (Loh & Elser, 2009, Ayyer *et al.*, 2016). Below we present a detailed description of all steps, which allowed us to obtain the electron density of PR772 virus with an improved resolution as compared to the previous SPI studies on the same virus (Rose *et al.,* 2018).

## 2. Experiment

The experiment was performed at the Atomic, Molecular and Optical Science (AMO) instrument (Bozek, 2009; Ferguson *et al.*, 2015; Osipov *et al.*, 2018) at the Linac Coherent Light

Source (LCLS) at SLAC National Accelerator Laboratory using the LAMP end station (see for experimental details (Li *et al.,* 2020)). PR772 bacteriophage growth and purification were done essentially as previously described (Reddy *et al.*, 2017). Viruses in ammonium acetate volatile buffer were aerosolized in a helium environment using a gas dynamic virtual nozzles (GDVN) that were 3D printed via 2-photon polymerization photo-lithography with a Nanoscribe Professional GT printer (Nazari *et al.,*2020). Laboratory measurements showed reproducible jet diameters in the range of 0.5-2.0 µm. (DePonte *et al.*, 2008; Weierstall *et al.*, 2012). The particles then passed through a differentially pumped skimmer for pressure reduction and were then injected/focused into the sample chamber using an aerodynamic lens injector (Benner *et al.*, 2008; Hantke *et al.*, 2014). The focused particle stream intersected the focused and pulsed XFEL beam.

The LCLS had a repetition rate of 120 Hz, for this experiment the average pulse energy was ~4 mJ, with a focal diameter of ~1.5 µm, and a photon energy of 1.7 keV (wavelength 0.729 nm). Diffraction patterns were recorded by a pnCCD detector (Strüder *et al*., 2010) mounted at a 0.130 m distance from the interaction region. In the experiment, a silver behenate salt was used as a calibration agent to determine sample detector distance and panel position. The scattering signal was recorded only by one of the two detector panels (one panel was not operational during the experiment due to an electronic fault). The size of the working panel was 512 by 1024 pixels with a pixel size of $75 \times 75$ µm$^2$, with the long edge closest to the interaction point. In the middle of the experiment (at run 205) the detector panel was moved one millimeter up vertically relative to the incoming X-ray beam to reduce background scattering. Our work is based on the data obtained from this panel covering part of reciprocal space as shown in Fig. 1.

### 3. Initial classification steps

SPI data analysis involves many subsequent steps, leading to the 3D reconstructed particle structure from a large set of 2D diffraction patterns (Gaffney & Chapman, 2007). Improvements in the data analysis pipeline at early stages can result in significant enhancement in reconstruction quality. Therefore, several pre-processing methods were developed to avoid experimental artifacts on the collected diffraction patterns. Pre-processing stages include: hit finding, background correction, beam position refinement, and particle size filtering.

The initial experimental dataset, as collected at the AMO instrument, consists of about $1.2 \times 10^7$ diffraction patterns ($9 \times 10^6$ patterns before and $3 \times 10^6$ patterns after moving the detector panel) (see (Li *et al.*, 2020)). The hit finding was performed using the software "psocake" in the "psana" framework (Damiani *et al.*, 2016). As a result, $1.9 \times 10^5$ diffraction patterns were identified as hits from the initial set of diffraction patterns, and the signal from these hits was converted to photon counts (see Table 1). Examples of selected diffraction patterns are shown in Fig. 1.

Visual inspection of the selected patterns revealed the presence of additional instrumental scattering near the center of the diffraction patterns. The scattering was caused by the interaction of the tails of the FEL beam with the upstream apertures or from the sample injector. To subtract this additional signal, histograms of intensity for each pixel were analyzed (see Supporting Information Fig. S1 and Fig. S2). It was assumed that for pixels with zero photon counts, the histogram of intensity distribution should have a Gaussian form and mean value equal to zero. The mean value of this distribution was subtracted from each diffraction pattern. Subtraction of this additional scattering contribution was an important step prior to the finding the beam center position (see Supporting Information for details).

Next, the particle size filtering was performed in two steps. First, the power spectral density (PSD) function, *i.e.* angular averaged intensity, of each diffraction pattern was fitted with the form-factor of a solid sphere in a wide range of sizes from 30 nm to 300 nm. Next, a fidelity criterion was introduced to distinguish between successful and unsuccessful size findings (see Supporting Information for details). As an outcome of this analysis, the number of selected diffraction patterns was reduced from $1.9 \times 10^5$ to $1.8 \times 10^5$, which were used for the final particle size filtering. A histogram for different particle sizes that ranges for the selected values is shown in Fig. 2. An extended range of sizes observed in this figure corresponds to clusters of particles stuck together and a varying thickness of a hydration layer. One can also identify in this histogram a peak in the range from 55 nm to 84 nm. This size range agrees well with the expected virus particle size of about 70 nm (Rose *et al.*, 2018; Reddy *et al.*, 2019). The considered range contains $1.8 \times 10^4$ diffraction patterns, which were selected for further single hit classification (see Table 1).

## 4. Single hit diffraction patterns classification

The key step of data selection in this work was single-hit classification. The angular X-ray cross-correlation analysis (AXCCA) classification that was used in the previous work (Bobkov *et al.*, 2015; Rose *et al.*, 2018) was not as effective with the present experimental data due to the absence of the scattering signal on one half of the detector and the low single-hit rate.

The single-hit classification approach was based on the expectation-maximization (EM) algorithm (Dempster *et al.*, 1977; Loh & Elser, 2009; Ayyer *et al.*, 2016). This algorithm allows for unsupervised clustering of data when neither initial data assignments to clusters nor cluster parameters are known. The data set is distributed into a pre-defined number of clusters and cluster parameters are retrieved automatically at the same time. Later, manual input is used to classify each of the cluster models, based on symmetry considerations or expected fringe patterns, in order to perform data selection. This algorithm is commonly used in Cryo-EM for unsupervised single-particle clustering (see, for example, software RELION (Scheres *et al.*, 2005)). As soon as, low contrast and low signal to noise level are common problems for Cryo-EM and SPI, we implemented in this work the original EM algorithm developed in Cryo-EM for clustering of SPI data.

Unsupervised EM-clustering is an iterative process; the algorithm starts from a random model for each class. At each iteration the probability for each diffraction pattern to belong to a certain class is calculated. To accommodate for random particle orientations in the SPI experiment, the cluster model is compared with the 2D diffraction pattern rotated in-plane. After probabilities are evaluated, a new cluster model is calculated by weighted averaging of patterns that belong to each cluster in each orientation. The weights are defined by computed probabilities. In fact, the algorithm imitates the EMC algorithm (Loh & Elser, 2009) but, instead of rotation in 3D space, an in-plane rotation and cluster distribution is used (Loh, 2014). When the EM-algorithm converges, the supervised class selection takes place. The cluster models that correspond to single-hits of an investigated particle are selected manually by an expert. As an example, in one of the EM-clustering we selected classes 1 and 2 as classes containing high contrast six fringes that we attribute to scattering from a PR772 virus particle (see Fig. 3(a)).

To make classification more accurate we performed five independent EM-clusterings starting from random cluster models. Each EM-clustering produced slightly different results, which are summarized in Table 1. The intersection of all results was considered as a stable single hit selection (see Fig. 3(b)). Finally, we ended with the data set containing 1,085 single particle patterns (see Table 1). The PSD function of an average of all selected diffraction patterns is shown in Fig. 3(c). The scattering signal from the virus particle extends up to momentum transfer value $q_{max}$=1 nm$^{-1}$ in reciprocal space, which corresponds to a resolution ($2\pi/q_{max}$) of 6.3 nm in real space.

A manual search and selection of single hits from the data set of $1.9\times10^5$ diffraction patterns produced a new data set containing 1,393 diffraction patterns (Li *et al.*, 2020). From this selection 574 diffraction patterns are also present in our EM-based selection. The PSD function for the manual selection shown in Fig. 3(c) has lower contrast without visible fringes. We attribute this mostly to the absence of the size filtering step in manual selection. The virus size fluctuations could be caused by a slight change in hydration layer thickness, variation of the position where the particles interact with X-ray beam, slightly upstream or downstream of the focus or a real sample size distribution of the PR772. We note that in our previous work (Rose *et al.*, 2018) the number of single hit diffraction patterns was about $7.3\times10^3$ which is about one order of magnitude higher than in this work. The reason for the smaller number of patterns was a combination of downtime due to detector troubleshooting and time needed for sample-injection optimizations. The GDVN flow and pressures needed to be optimized because of the relatively large initial droplet diameters of approximately one to two micrometers.

## 5. Orientation determination and background subtraction

For orientation determination we used the Expand–Maximize–Compress (EMC) algorithm implemented in the Dragonfly software package (Ayyer *et al.*, 2016). This iterative algorithm successfully combines 2D diffraction patterns into 3D intensity distribution of the PR772 virus shown in Fig. 4(a) (see Supporting Information for further details).

Results presented in Fig. 4(a) clearly show that the recovered 3D diffraction intensity contains a substantial high momentum background that may be caused by scattering on helium from the carrier gas. To reduce the influence of this background contribution, a background subtraction was applied. Several approaches for the background correction in SPI experiment analysis were developed earlier (Rose *et al.*, 2018; Lundholm *et al.*, 2018; Ayyer et al., 2019) and it was understood that the background correction method may affect the reconstructed structure.

Here we used a combined approach for the background determination, first a constant background was subtracted and then, on the reconstruction step, the contrast of diffraction patterns was additionally enhanced by applying deconvolution algorithms (see next section for details). In the first step, the background level was defined as the mean signal value in the high momentum region of the 3D intensity distribution free of particle scattering contribution (see Supporting Information for details). Reciprocal space data with and without background subtraction are shown in Fig. 4(a-c). As it is seen in Fig. 4(c), the power law dependence of data after the background subtraction is the same up to high $q$-range of about one inverse nanometer. That is in contrast to the data after EMC orientation determination (blue curve in Fig. 4(c)), which is clearly saturated at high $q$-range.

Due to the background subtraction, more features and higher contrast were revealed in the high momentum transfer region. The fringe visibility or averaged contrast $\langle \gamma \rangle$ was defined as

$$\langle \gamma \rangle = \frac{1}{N}\sum_{i=1}^{N} \gamma_i \text{, where } \gamma_i = \frac{I_{max}^i - I_{min}^i}{I_{max}^i + I_{min}^i}. \tag{1}$$

Here $\gamma_i$ is the local contrast and $I_{max}^i$ and $I_{min}^i$ are the PSD-function values in the local maxima and following minima, respectively and $N$ is a number of pairs of maxima and minima considered in this analysis. In our case the average contrast values $\langle \gamma \rangle$ were calculated for the first $N$=6 pairs. For the experimental data before the background subtraction we obtained the value $\langle \gamma \rangle$=0.41. The fringe contrast after the background subtraction showed significant improvement with $\langle \gamma \rangle$=0.58.

To proceed further with the analysis, low ($q < 0.12$ nm$^{-1}$) and high ($q > 0.93$ nm$^{-1}$) momentum transfer regions were eliminated from the final data selection due to scattering artifacts present in these regions. After the background subtraction, negative values of intensity were set to zero. A 3D intensity distribution in reciprocal space after the background subtraction is shown in Fig. 4(d).

## 6. Phase retrieval and PR772 virus structure

To determine the electron density distribution of the PR772 virus the phase retrieval was performed using the 3D intensity distribution of the virus in reciprocal space described in the previous section (see Fig. 4(d)). Iterative phase retrieval algorithms are based on the Fourier transform between real and reciprocal/diffraction space using two constraints: in reciprocal space the amplitude of the signal is set equal to the experimentally measured values and in real space a finite support of the particle is used (Fienup, 1982; Marchesini, 2007).

The electron density distribution of the PR772 virus was obtained with the following steps. First, the central gap, originating from the initial data masking of the diffraction patterns, was determined (see Fig. 4(a,b)). This was accomplished by running multiple 3D reconstructions of the virus and leaving the central gap of the diffraction volume to freely evolve. All reconstructions produced practically identical continuous 3D functions in reciprocal space that smoothly covered the central gap (see Fig. S10 in Supplementary Information). This reciprocal space distribution with the filled central part was used for the next reconstruction step. On that step 50 successive reconstructions using a combination of continuous hybrid input–output (cHIO) (Fienup, 2013), error reduction (ER) (Fienup, 1982), and shrink-wrap (SW) approach (Marchesini *et al.*, 2003) were performed. In addition to these algorithms, we used the Richardson-Lucy deconvolution technique (Clark *et al.*, 2012). Application of this method allowed to take into consideration additional background and substantially enhance the contrast of the reconstructed diffraction pattern to the value of $\langle \gamma \rangle = 0.87$. As a result, we obtained complex valued real space images for each 50 reconstructions (see Supporting Information for details).

On the final step, to identify electron density of the virus we used mode decomposition. We first introduce density matrix in the form

$$\rho(r_1, r_2) = \langle \rho^*(r_1)\rho(r_2) \rangle, \tag{2}$$

where $\rho(r)$ are complex valued real space images and brackets $<...>$ indicate ensemble averaging over different reconstructions. We further decompose the density matrix into orthogonal modes $\rho_n(r)$ using Principal Component Analysis (PCA) (Khubbutdinov *et al., 2019)

$$\rho(r_1, r_2) = \sum_{n=0}^{N} \beta_n \rho_n^*(r_1)\rho_n(r_2), \tag{3}$$

where $\beta_n$ are eigenvalues of this decomposition. This approach is advantageous in comparison to averaging that would make important object features blurry and, finally, may affect the final resolution. By considering the zero mode only, unique features present in all other modes will be represented. In practice, we used all 50 reconstruction results and performed mode decomposition. The fundamental mode (with the weight factor $\beta_0$ of 99%) was considered as the final result of reconstruction. Such a high weight factor value indicates that all reconstructions converged essentially to the same result with the uncertainty level of only 1%. To determine the electron density, we took an absolute value of this fundamental mode complex valued image. Three-times up-sampled version of the real space electron density is shown in Fig. 5 (see Supporting Information for the details).

As is seen in Fig. 5(a-c), the retrieved electron density of the PR772 virus shows the expected icosahedral structure. A 2D cut of the virus structure is presented in Fig. 5(d), where a higher electron density in a thin outer shell is well resolved and is attributed to the capsid proteins arrangement. As one can see in Fig. 5(d) the electron density in the center of the reconstructed virus particle was reduced. The reason for this may be heterogeneity of virus particles present in solution and injected in X-ray beam.

PR772, like other members of the *Tectiviridae* family, contains an inner proteolipid membrane that facilitates delivery of the viral genomic DNA during infection (Mantynen et al., 2019). When the virus binds to a bacterial host cell the inner membrane is extruded from one of the viral vertices to form a nanotube that facilitates genome delivery. Spontaneous release of the viral genome has been reported (Peralta *et al.,* 2013; Santos-Perez *et al.,* 2017). Our PR772 preparations were also analyzed by Cryo-EM (see Supporting Information Fig. S6). Under the conditions used for plunge freezing, we observed particles that were more rounded than icosahedral. Preliminary volume analysis suggests that some particles had released their DNA. These particles appeared to be "triggered" during the freezing process since transmission electron microscopy (TEM) imaging did not reveal this. It may be possible that minor differences in virus preparation or upon XFEL sample delivery that some of the virus particles were similarly "triggered" to release their genomes during cryo-freezing, thus resulting in decrease in inner electron density. Previous analyses of PR772 XFEL snapshots also suggested that some particles exhibit decreased inner density (Shi *et al.,* 2019; Hosseinizadeh *et al.,* 2017).

To identify the particle size the electron density profiles in different directions were investigated. In this work, we used the same approach as developed earlier (Rose *et al.*, 2018) and analyzed the electron density profiles in the directions from facet to facet and from vertex to vertex of the reconstructed virus particle (see Fig. 6(a,b)). For the particle size estimate we selected the electron density threshold value of 0.2 as it was considered in the shrink-wrap algorithm during the phase retrieval. From this criterion we determined the particle size in different directions (see Table 2 and Supporting Information). Thus, the obtained mean particle size between facets was 61±2 nm and between vertices 63±2 nm, respectively. These sizes correspond well to the initial range of particle sizes (from 55 nm to 84 nm) considered at the initial classification step (see section IV) and other XFEL data performed on the same virus (Kurta et al., 2017; Rose et al., 2018).

Similar to previous work (Kurta *et al.*, 2017; Rose *et al.*, 2018) we observed a certain elongation of the particle shape, which might be inherently present in the viruses in solution or appear due to the aerosol injection conditions. We defined elongation of the particle by the following measure

$$\alpha = \frac{S_{max} - S_{min}}{S_{mean}}, \tag{4}$$

where $S_{max}$, $S_{min}$, and $S_{mean}$ are maximum, minimum, and mean particle sizes, respectively. The virus structure obtained in the current work showed elongation value $\alpha$=11% for sizes taken between vertices, which is similar to the result of the previous SPI experiments (Rose *et al.*, 2018) with $\alpha$=9%.

The mean capsid thickness was obtained from the Gaussian fitting of the electron density profiles with well pronounced features (see Supporting Information Fig. S14) and was determined to be 7.6 ± 0.3 nm. The thickness of the capsid in recent Cryo-EM studies (Reddy *et al.*, 2019) was identified to be below 10 nm, which is in good agreement with the thickness determined in this experiment.

Finally, we determined resolution of the reconstructed electron density of the PR772 virus. We used the Fourier Shell Correlation (FSC) approach (Harauz & van Heel, 1986) to determine resolution of the reconstructed virus. For this method two independent sets of reconstruction are required, usually each is based on half of the available data set. FSC measures the normalized cross-correlation coefficient between both reconstructions over corresponding shells in Fourier space. Results of the FSC analysis are shown in Fig. 7. To estimate the achieved resolution we used ½-bit threshold criteria (van Heel & Schatz, 2005). Its intersection with the FSC-curve gave a resolution value of 6.9 nm (see Fig. 7). The obtained result is better than previously reported value of ~8 nm for the same virus (Rose *et al.*, 2018), though the number of diffraction patterns used for the final analysis was much lower (20% of the previous data set). In this case the resolution is limited by the number of diffraction patterns and moderate scattered intensity where previous reconstructions were limited by the extent of the detector.

### 7. Summary and Outlook

We have presented implementation of the SPI data analysis workflow from diffraction patterns measured at the AMO instrument at LCLS to reconstruct the electron density of the PR772 virus. This work has an important methodological step for development of the SPI data analysis. We implemented several methods into the workflow including EM-based classification and mode decomposition that were crucial for the high-resolution final reconstruction. Although only half of the detector was operational, implementation of all these steps allowed to determine the PR772 virus structure with a higher resolution as compared to previous SPI studies.

From the initial set of $1.2 \times 10^7$ experimentally measured diffraction patterns, the final single hit selection set contained 1,085 diffraction patterns. About 53% of this final set was also present in the single hit selection made manually. The number of diffraction patterns classified for further analysis was substantially lower than in the previous experiment (Rose *et al.*, 2018), improvements in sample delivery to increase the number of diffraction patterns is critical for improvements in reconstruction resolution.

The combination of all methods implemented in the workflow allowed to obtain the 3D electron density of the virus with the resolution of 6.9 nm based on the FSC analysis, that is better than obtained in the previous studies (Rose *et al.*, 2018). The obtained mean PR772 virus size in this experiment was 61 nm (between facets) and 63 nm (between vertices), that is on the same order as in the previous SPI experiments performed on the same PR772 virus (Kurta *et al.*, 2017; Rose *et al.*, 2018). We also observed a similar elongation of about 11% of the virus structure as it was determined in the previous experiments.

The presented research is another step forward in the SPI data analysis. Implemented steps may become especially important when SPI experiments will be performed at high repetition rate XFELs such as the European XFEL (Decking *et al*., 2020; Mancuso *et al*., 2019) and LCLS-II (Schoenlein *et al.*, 2015) facilities. The first experiments performed at the European XFEL demonstrated the possibility of collecting the SPI data at megahertz rate (Sobolev *et al.*, 2019) that might be crucial for the future progress in single particle imaging experiments performed at XFELs.

**Funding**

DA, YYK, SB, RK, MR, AT, VAI, IAV acknowledge support from the Helmholtz Associations Initiative Networking Fund and the Russian Science Foundation grant HRSF-0002/18-41-06001. HDM acknowledges US National Science Foundation (NSF) Science and Technology Center BioXFEL Award 1231306. JA acknowledges support from the project "Advanced research using high intensity laser produced photons and particles" (CZ.02.1.01/0.0/0.0/16_019/0000789) from European Regional Development Fund (ADONIS) and the Ministry of Education, Youth and Sports as part of targeted support from the National Program of Sustainability II. PLX acknowledges Grant: European Research Council, "Frontiers in Attosecond X-ray Science: Imaging and Spectroscopy (AXSIS)", ERC-2013-SyG 609920 (2014-2018), the Human Frontier Science Program (RGP0010/2017), and the fellowship from the Joachim Herz Stiftung. JH acknowledges the Swedish Research Council, the Knut and Alice Wallenberg Foundation, the European Research Council, the Röntgen–Ångström Cluster, and by the project "Structural dynamics of biomolecular systems" (CZ.02.1.01/0.0/0.0/15_003/0000447) from the European Regional Development Fund. EB acknowledges support by the Australian Research Council Centre of Excellence in Advanced Molecular Imaging (CE140100011).

**Acknowledgements**

The authors acknowledge use of the Linac Coherent Light Source (LCLS), SLAC National Accelerator Laboratory, use is supported by the U.S. Department of Energy, Office of Science, Office of Basic Energy Sciences under Contract No. DE-AC02-76SF00515. The authors also acknowledge the invaluable support of the technical and scientific staff of the LCLS at SLAC National Accelerator Laboratory. The authors also acknowledge support of the project and discussions with E. Weckert and careful reading of the manuscript by A. Schropp.

**References**

Allahgholi, A., Becker, J., Delfs, A., Dinapoli, R., Goettlicher, P., Greiffenberg, D., Henrich, B., Hirsemann, H., Kuhn, M., Klanner, R., Klyuev, A., Krueger, H., Lange, S., Laurus, T., Marras, A., Mezza, D., Mozzanica, A., Niemann, M., Poehlsen, J., Schwandt, J., Sheviakov, I., Shi, X., Smoljanin, S., Steffen, L., Sztuk-Dambietz, J., Trunk, U., Xia, Q., Zeribi, M., Zhang, J., Zimmer, M., Schmitt, B. & Graafsma, H. (2019). *J. Synchrotron Rad.* **26**, 74–82.

Aquila, A., Barty, A., Bostedt, C., Boutet, S., Carini, G., DePonte, D., Drell, P., Doniach, S., Downing, K. H., Earnest, T., Elmlund, H., Elser, V., Gühr, M., Hajdu, J., Hastings, J., Hau-Riege, S. P., Huang, Z., Lattman, E. E., Maia, F. R. N. C., Marchesini, S., Ourmazd, A., Pellegrini, C., Santra, R., Schlichting, I., Schroer, C., Spence, J. C. H., Vartanyants, I. A., Wakatsuki, S., Weis, W. I. & Williams, G. J. (2015). *Struct. Dyn.* **2**, 041701.

Aquila, A., Hunter, M. S., Doak, R. B., Kirian, R. A., Fromme, P., White, T. A., Andreasson, J., Arnlund, D., Bajt, S., Barends, T. R. M., Barthelmess, M., Bogan, M. J., Bostedt, C., Bottin, H., Bozek, J. D., Caleman, C., Coppola, N., Davidsson, J., DePonte, D. P., Elser, V., Epp, S. W., Erk, B., Fleckenstein, H., Foucar, L., Frank, M., Fromme, R., Graafsma, H., Grotjohann, I., Gumprecht, L., Hajdu, J., Hampton, C.

Y., Hartmann, A., Hartmann, R., Hau-Riege, S., Hauser, G., Hirsemann, H., Holl, P., Holton, J. M., Hömke, A., Johansson, L., Kimmel, N., Kassemeyer, S., Krasniqi, F., Kühnel, K.-U., Liang, M., Lomb, L., Malmberg, E., Marchesini, S., Martin, A. V., Maia, F. R. N. C., Messerschmidt, M., Nass, K., Reich, C., Neutze, R., Rolles, D., Rudek, B., Rudenko, A., Schlichting, I., Schmidt, C., Schmidt, K. E., Schulz, J., Seibert, M. M., Shoeman, R. L., Sierra, R., Soltau, H., Starodub, D., Stellato, F., Stern, S., Strüder, L., Timneanu, N., Ullrich, J., Wang, X., Williams, G. J., Weidenspointner, G., Weierstall, U., Wunderer, C., Barty, A., Spence, J. C. H. & Chapman, H. N. (2012). *Opt. Express.* **20**, 2706.

Ayyer, K., Lan, T.-Y., Elser, V. & Loh, N. D. (2016). *J. Appl. Crystallogr.* **49**, 1320–1335.

Ayyer, K., Morgan, A. J., Aquila, A., DeMirci, H., Hogue, B. G., Kirian, R. A., Xavier, P. L., Yoon, C. H., Chapman, H. N. & Barty, A. (2019). *Opt. Express.* **27**, 37816.

Benner, W. H., Bogan, M. J., Rohner, U., Boutet, S., Woods, B. & Frank, M. (2008). *J. Aerosol Sci.* **39**, 917–928.

Bobkov, S. A., Teslyuk, A. B., Kurta, R. P., Gorobtsov, O. Y., Yefanov, O. M., Ilyin, V. A., Senin, R. A. & Vartanyants, I. A. (2015). *J. Synchrotron Rad.* **22**, 1345–1352.

Bogan, M. J., Benner, W. H., Boutet, S., Rohner, U., Frank, M., Barty, A., Seibert, M. M., Maia, F., Marchesini, S., Bajt, S., Woods, B., Riot, V., Hau-Riege, S. P., Svenda, M., Marklund, E., Spiller, E., Hajdu, J., & Chapman, H. N. (2008). *Nano Lett.,* 8, 310-316.

Boutet, S., Lomb, L., Williams, G. J., Barends, T. R. M., Aquila, A., Doak, R. B., Weierstall, U., DePonte, D. P., Steinbrener, J., Shoeman, R. L., Messerschmidt, M., Barty, A., White, T. A., Kassemeyer, S., Kirian, R. A., Seibert, M. M., Montanez, P. A., Kenney, C., Herbst, R., Hart, P., Pines, J., Haller, G., Gruner, S. M., Philipp, H. T., Tate, M. W., Hromalik, M., Koerner, L. J., van Bakel, N., Morse, J., Ghonsalves, W., Arnlund, D., Bogan, M. J., Caleman, C., Fromme, R., Hampton, C. Y., Hunter, M. S., Johansson, L. C., Katona, G., Kupitz, C., Liang, M., Martin, A. V., Nass, K., Redecke, L., Stellato, F., Timneanu, N., Wang, D., Zatsepin, N. A., Schafer, D., Defever, J., Neutze, R., Fromme, P., Spence, J. C. H., Chapman, H. N. & Schlichting, I. (2012). *Science* **337**, 362–364.

Bozek, J. D. (2009). *Eur. Phys. J. Spec. Top.* **169**, 129–132.

Chapman, H. N., Barty, A., Bogan, M. J., Boutet, S., Frank, M., Hau-Riege, S. P., Marchesini, S., Woods, B. W., Bajt, S., Benner, W. H., London, R. A., Plönjes, E., Kuhlmann, M., Treusch, R., Düsterer, S., Tschentscher, T., Schneider, J. R., Spiller, E., Möller, T., Bostedt, C., Hoener, M., Shapiro, D. A., Hodgson, K. O., van der Spoel, D., Burmeister, F., Bergh, M., Caleman, C., Huldt, G., Seibert, M. M., Maia, F. R. N. C., Lee, R. W., Szöke, A., Timneanu, N. & Hajdu, J. (2006). *Nat. Phys.* **2**, 839–843.

Chapman, H. N., Fromme, P., Barty, A., White, T. A., Kirian, R. A., Aquila, A., Hunter, M. S., Schulz, J., DePonte, D. P., Weierstall, U., Doak, R. B., Maia, F. R. N. C., Martin, A. V., Schlichting, I., Lomb, L., Coppola, N., Shoeman, R. L., Epp, S. W., Hartmann, R., Rolles, D., Rudenko, A., Foucar, L., Kimmel, N., Weidenspointner, G., Holl, P., Liang, M., Barthelmess, M., Caleman, C., Boutet, S., Bogan, M. J., Krzywinski, J., Bostedt, C., Bajt, S., Gumprecht, L., Rudek, B., Erk, B., Schmidt, C., Hömke, A., Reich, C., Pietschner, D., Strüder, L., Hauser, G., Gorke, H., Ullrich, J., Herrmann, S., Schaller, G., Schopper, F., Soltau, H., Kühnel, K.-U., Messerschmidt, M., Bozek, J. D., Hau-Riege, S. P., Frank, M., Hampton, C. Y., Sierra, R. G., Starodub, D., Williams, G. J., Hajdu, J., Timneanu, N., Seibert, M. M., Andreasson, J., Rocker, A., Jönsson, O., Svenda, M., Stern, S., Nass, K., Andritschke, R., Schröter, C.-D., Krasniqi, F., Bott, M., Schmidt, K. E., Wang, X., Grotjohann, I., Holton, J. M., Barends, T. R. M., Neutze, R., Marchesini, S., Fromme, R., Schorb, S., Rupp, D., Adolph, M., Gorkhover, T., Andersson, I., Hirsemann, H., Potdevin, G., Graafsma, H., Nilsson, B. & Spence, J. C. H. (2011). *Nature.* **470**, 73–77.

Clark, J. N., Huang, X., Harder, R. & Robinson, I. K. (2012). *Nat. Commun.* **3**, 993.

Damiani, D., Dubrovin, M., Gaponenko, I., Kroeger, W., Lane, T. J., Mitra, A., O'Grady, C. P., Salnikov, A., Sanchez-Gonzalez, A., Schneider, D. & Yoon, C. H. (2016). *J. Appl. Crystallogr.* **49**, 672–679.

Decking, W. *et al*. (2020). *Nat. Photon.* **14**, 391–397.

Dempster, A. P., Laird, N. M. & Rubin, D. B. (1977). *J. R. Stat. Soc. Ser. B*. **39**, 1–22.

DePonte, D. P., Weierstall, U., Schmidt, K., Warner, J., Starodub, D., Spence, J. C. H. & Doak, R. B. (2008). *J. Phys. D. Appl. Phys.* **41**, 195505.

Ekeberg, T., Svenda, M., Abergel, C., Maia, F. R. N. C., Seltzer, V., Claverie, J.-M., Hantke, M., Jönsson, O., Nettelblad, C., van der Schot, G., Liang, M., DePonte, D. P., Barty, A., Seibert, M. M., Iwan, B., Andersson, I., Loh, N. D., Martin, A. V., Chapman, H., Bostedt, C., Bozek, J. D., Ferguson, K. R., Krzywinski, J., Epp, S. W., Rolles, D., Rudenko, A., Hartmann, R., Kimmel, N. & Hajdu, J. (2015). *Phys. Rev. Lett.* **114**, 098102.

Emma, P., Akre, R., Arthur, J., Bionta, R., Bostedt, C., Bozek, J., Brachmann, A., Bucksbaum, P., Coffee, R., Decker, F.-J., Ding, Y., Dowell, D., Edstrom, S., Fisher, A., Frisch, J., Gilevich, S., Hastings, J., Hays, G., Hering, P., Huang, Z., Iverson, R., Loos, H., Messerschmidt, M., Miahnahri, A., Moeller, S., Nuhn, H.-D., Pile, G., Ratner, D., Rzepiela, J., Schultz, D., Smith, T., Stefan, P., Tompkins, H., Turner, J., Welch, J., White, W., Wu, J., Yocky, G. & Galayda, J. (2010). *Nat. Photon.* **4**, 641–647.

Ferguson, K. R., Bucher, M., Bozek, J. D., Carron, S., Castagna, J.-C., Coffee, R., Curiel, G. I., Holmes, M., Krzywinski, J., Messerschmidt, M., Minitti, M., Mitra, A., Moeller, S., Noonan, P., Osipov, T., Schorb, S., Swiggers, M., Wallace, A., Yin, J. & Bostedt, C. (2015). *J. Synchrotron Rad.* **22**, 492–497.

Fienup, J. R. (1982). *Appl. Opt.* **21**, 2758.

Fienup, J. R. (2013). *Appl. Opt.* **52**, 45.

Gaffney, K. J. & Chapman, H. N. (2007). *Science* **316**, 1444–1448.

Gorobtsov, O. Y., Lorenz, U., Kabachnik, N. M. & Vartanyants, I. A. (2015). *Phys. Rev. E.* **91**, 062712.

Hantke, M. F., Hasse, D., Maia, F. R. N. C., Ekeberg, T., John, K., Svenda, M., Loh, N. D., Martin, A. V, Timneanu, N., Larsson, D. S. D., van der Schot, G., Carlsson, G. H., Ingelman, M., Andreasson, J., Westphal, D., Liang, M., Stellato, F., DePonte, D. P., Hartmann, R., Kimmel, N., Kirian, R. A., Seibert, M. M., Mühlig, K., Schorb, S., Ferguson, K., Bostedt, C., Carron, S., Bozek, J. D., Rolles, D., Rudenko, A., Epp, S., Chapman, H. N., Barty, A., Hajdu, J. & Andersson, I. (2014). *Nat. Photon.* **8**, 943–949.

Harauz, G. & van Heel, M. (1986). *Optik* **73**, 146–156.

van Heel, M. & Schatz, M. (2005). *J. Struct. Biol.* **151**, 250–262.

Hosseinizadeh, A., Mashayekhi, G., Copperman, J., Schwander, P., Dashti, A., Sepehr, R., Fung, R., Schmidt, M., Yoon, C.H., Hogue, B.G., Williams, G.J., Aquila, A., Ourmazd, A. (2017) *Nat. Meth.* **14,** 877-881.

Ishikawa, T., Aoyagi, H., Asaka, T., Asano, Y., Azumi, N., Bizen, T., Ego, H., Fukami, K., Fukui, T., Furukawa, Y., Goto, S., Hanaki, H., Hara, T., Hasegawa, T., Hatsui, T., Higashiya, A., Hirono, T., Hosoda, N., Ishii, M., Inagaki, T., Inubushi, Y., Itoga, T., Joti, Y., Kago, M., Kameshima, T., Kimura, H., Kirihara, Y., Kiyomichi, A., Kobayashi, T., Kondo, C., Kudo, T., Maesaka, H., Maréchal, X. M., Masuda, T., Matsubara, S., Matsumoto, T., Matsushita, T., Matsui, S., Nagasono, M., Nariyama, N., Ohashi, H., Ohata, T., Ohshima, T., Ono, S., Otake, Y., Saji, C., Sakurai, T., Sato, T., Sawada, K., Seike, T., Shirasawa, K., Sugimoto, T., Suzuki, S., Takahashi, S., Takebe, H., Takeshita, K., Tamasaku, K., Tanaka, H., Tanaka, R., Tanaka, T., Togashi, T., Togawa, K., Tokuhisa, A., Tomizawa, H., Tono, K., Wu, S., Yabashi, M., Yamaga, M., Yamashita, A., Yanagida, K., Zhang, C., Shintake, T., Kitamura, H. & Kumagai, N. (2012). *Nat. Photon.* **6**, 540–544.

Khubbutdinov, R., Menushenkov, A. P. & Vartanyants, I. A. (2019). *J. Synchrotron Rad.* **26**, 1851–1862.

Kurta, R. P., Donatelli, J. J., Yoon, C. H., Berntsen, P., Bielecki, J., Daurer, B. J., DeMirci, H., Fromme, P., Hantke, M. F., Maia, F. R. N. C., Munke, A., Nettelblad, C., Pande, K., Reddy, H. K. N., Sellberg, J. A., Sierra, R. G., Svenda, M., van der Schot, G., Vartanyants, I. A., Williams, G. J., Xavier, P. L., Aquila, A., Zwart, P. H. & Mancuso, A. P. (2017). *Phys. Rev. Lett.* **119**, 158102.

Li, H., Aquila, A., Abbey, B., Alvarez, R., Ayyer, K., Barty, A., Berntsen, P., Bielecki, J., Bucher, M., Carini, G., Chapman, H. N., Contreras, A., Daurer, B. J., DeMirci, H., Flückiger, L., Frank, M., Hogue, B. G., Hosseinizadeh, A., Jönsson, H. O., Kirian, R. A., Kurta, R. P., Loh, D., Maia, F. R. N. C., Mancuso, A., McFadden, M., Muehlig, K., Munke A., Narayana Reddy, H. K., Nazari, R., Ourmazd, A., Rose, M., Schwander, P., Seibert, M. M., Sellberg, J. A., Sierra, R. G., Sun, Z., Svenda, M., Vartaniants, I., Walter, P., Williams, G., Xavier, P., L., Zaare, S. (2020). *Sci. Data.* (submitted).

Loh, N.-T. D. & Elser, V. (2009). *Phys. Rev. E.* **80**, 026705.

Loh, N.-T. D. (2014). *Philos. Trans. R. Soc. B Biol. Sci.* **369**, 20130328.

Lorenz, U., Kabachnik, N. M., Weckert, E. & Vartanyants, I. A. (2012). *Phys. Rev. E.* **86**, 051911.

Lundholm, I. V, Sellberg, J. A., Ekeberg, T., Hantke, M. F., Okamoto, K., van der Schot, G., Andreasson, J., Barty, A., Bielecki, J., Bruza, P., Bucher, M., Carron, S., Daurer, B. J., Ferguson, K., Hasse, D., Krzywinski, J., Larsson, D. S. D., Morgan, A., Mühlig, K., Müller, M., Nettelblad, C., Pietrini, A., Reddy, H. K. N., Rupp, D., Sauppe, M., Seibert, M., Svenda, M., Swiggers, M., Timneanu, N., Ulmer, A., Westphal, D., Williams, G., Zani, A., Faigel, G., Chapman, H. N., Möller, T., Bostedt, C., Hajdu, J., Gorkhover, T. & Maia, F. R. N. C. (2018). *IUCrJ.* **5**, 531–541.

Mancuso, A. P., Aquila, A., Batchelor, L., Bean, R. J., Bielecki, J., Borchers, G., Doerner, K., Giewekemeyer, K., Graceffa, R., Kelsey, O. D., Kim, Y., Kirkwood, H. J., Legrand, A., Letrun, R., Manning, B., Lopez Morillo, L., Messerschmidt, M., Mills, G., Raabe, S., Reimers, N., Round, A., Sato, T., Schulz, J., Signe Takem, C., Sikorski, M., Stern, S., Thute, P., Vagovič, P., Weinhausen, B. & Tschentscher, T. (2019). *J. Synchrotron Rad.* **26**, 660–676.

Mäntynen, S., Sundberg, L.-R., Oksanen, H. M., and Poranen, M. M. (2019). *Viruses* **11**, 76.

Marchesini, S. (2007). *Rev. Sci. Instrum.* **78**, 011301.

Marchesini, S., He, H., Chapman, H. N., Hau-Riege, S. P., Noy, A., Howells, M. R., Weierstall, U. & Spence, J. C. H. (2003). *Phys. Rev. B.* **68**, 140101.

Munke, A., Andreasson, J., Aquila, A., Awel, S., Ayyer, K., Barty, A., Bean, R. J., Berntsen, P., Bielecki, J., Boutet, S., Bucher, M., Chapman, H. N., Daurer, B. J., DeMirci, H., Elser, V., Fromme, P., Hajdu, J., Hantke, M. F., Higashiura, A., Hogue, B. G., Hosseinizadeh, A., Kim, Y., Kirian, R. A., Reddy, H. K. N., Lan, T.-Y., Larsson, D. S. D., Liu, H., Loh, N. D., Maia, F. R. N. C., Mancuso, A. P., Mühlig, K., Nakagawa, A., Nam, D., Nelson, G., Nettelblad, C., Okamoto, K., Ourmazd, A., Rose, M., van der Schot, G., Schwander, P., Seibert, M. M., Sellberg, J. A., Sierra, R. G., Song, C., Svenda, M., Timneanu, N., Vartanyants, I. A., Westphal, D., Wiedorn, M. O., Williams, G. J., Xavier, P. L., Yoon, C. H. & Zook, J. (2016). *Sci. Data*. **3**, 160064.

Nazari, R., Zaare, S., Alvarez, R., , Karpos, K., Engelman, T., Madsen, C., Nelson, G., Spence J. C. H., Weierstall, U., Adrian, R. J., and Kirian, R. A. (2020) *Opt. Express* (in review).

Neutze, R., Wouts, R., van der Spoel, D., Weckert, E. & Hajdu, J. (2000). *Nature*. **406**, 752–757.

Osipov, T., Bostedt, C., Castagna, J.-C., Ferguson, K. R., Bucher, M., Montero, S. C., Swiggers, M. L., Obaid, R., Rolles, D., Rudenko, A., Bozek, J. D. & Berrah, N. (2018). *Rev. Sci. Instrum.* **89**, 035112.

Peralta, B., Gil-Carton, D., Castaño-Díez, D., Bertin, A., Boulogne, C., Oksanen, H. M., Bamford, D. H., and Abrescia, N. G. A. (2013). *PLoS Biol.*, **11**, e1001667.

Reddy, H. K. N., Carroni, M., Hajdu, J. & Svenda, M. (2019). *Elife*. **8**, e48496.

Reddy, H. K. N., Yoon, C. H., Aquila, A., Awel, S., Ayyer, K., Barty, A., Berntsen, P., Bielecki, J., Bobkov, S., Bucher, M., Carini, G. A., Carron, S., Chapman, H., Daurer, B., DeMirci, H., Ekeberg, T., Fromme, P., Hajdu, J., Hanke, M. F., Hart, P., Hogue, B. G., Hosseinizadeh, A., Kim, Y., Kirian, R. A., Kurta, R. P., Larsson, D. S. D., Duane Loh, N., Maia, F. R. N. C., Mancuso, A. P., Mühlig, K., Munke, A., Nam, D., Nettelblad, C., Ourmazd, A., Rose, M., Schwander, P., Seibert, M., Sellberg, J. A., Song, C., Spence, J. C. H., Svenda, M., Van der Schot, G., Vartanyants, I. A., Williams, G. J. & Xavier, P. L. (2017). *Sci. Data*. **4**, 170079.

Rose, M., Bobkov, S., Ayyer, K., Kurta, R. P., Dzhigaev, D., Kim, Y. Y., Morgan, A. J., Yoon, C. H., Westphal, D., Bielecki, J., Sellberg, J. A., Williams, G., Maia, F. R. N. C., Yefanov, O. M., Ilyin, V., Mancuso, A. P., Chapman, H. N., Hogue, B. G., Aquila, A., Barty, A. & Vartanyants, I. A. (2018). *IUCrJ*. **5**, 727–736.

Santos-Perez, I., Oksanen, H.M., Bamford, D. H., Goni, F.M., Reguera, D., Abrescia, N. G. A. (2017). *Biochim. Biophys. Acta* **1861**, 664-672.

Scheres, S. H. W., Valle, M., Nuñez, R., Sorzano, C. O. S., Marabini, R., Herman, G. T. & Carazo, J.-M. (2005). *J. Mol. Biol.* **348**, 139–149.

Schoenlein, R., Abbamonte, P., Abild-Pedersen, F., Adams, P., Ahmed, M., Albert, F., Mori, R. A., Anfinrud, A., Aquila, A. & Armstrong, M. (2015). *New Science Opportunities Enabled By LCLS-II X-Ray Lasers.* SLAC, Menlo Park, CA, USA, 1–189.

van der Schot, G., Svenda, M., Maia, F. R. N. C., Hantke, M., DePonte, D. P., Seibert, M. M., Aquila, A., Schulz, J., Kirian, R., Liang, M., Stellato, F., Iwan, B., Andreasson, J., Timneanu, N., Westphal, D., Almeida, F. N., Odic, D., Hasse, D., Carlsson, G. H., Larsson, D. S. D., Barty, A., Martin, A. V., Schorb, S., Bostedt, C., Bozek, J. D., Rolles, D., Rudenko, A., Epp, S., Foucar, L., Rudek, B., Hartmann, R., Kimmel, N., Holl, P., Englert, L., Duane Loh, N.-T., Chapman, H. N., Andersson, I., Hajdu, J. & Ekeberg, T. (2015). *Nat. Commun.* **6**, 5704.

Seibert, M. M., Ekeberg, T., Maia, F. R. N. C., Svenda, M., Andreasson, J., Jönsson, O., Odić, D., Iwan, B., Rocker, A., Westphal, D., Hantke, M., DePonte, D. P., Barty, A., Schulz, J., Gumprecht, L., Coppola, N., Aquila, A., Liang, M., White, T. A., Martin, A., Caleman, C., Stern, S., Abergel, C., Seltzer, V., Claverie, J.-M., Bostedt, C., Bozek, J. D., Boutet, S., Miahnahri, A. A., Messerschmidt, M., Krzywinski, J., Williams, G., Hodgson, K. O., Bogan, M. J., Hampton, C. Y., Sierra, R. G., Starodub, D., Andersson, I., Bajt, S., Barthelmess, M., Spence, J. C. H., Fromme, P., Weierstall, U., Kirian, R., Hunter, M., Doak, R. B., Marchesini, S., Hau-Riege, S. P., Frank, M., Shoeman, R. L., Lomb, L., Epp, S. W., Hartmann, R., Rolles, D., Rudenko, A., Schmidt, C., Foucar, L., Kimmel, N., Holl, P., Rudek, B., Erk, B., Hömke, A., Reich, C., Pietschner, D., Weidenspointner, G., Strüder, L., Hauser, G., Gorke, H., Ullrich, J., Schlichting, I., Herrmann, S., Schaller, G., Schopper, F., Soltau, H., Kühnel, K.-U., Andritschke, R., Schröter, C.-D., Krasniqi, F., Bott, M., Schorb, S., Rupp, D., Adolph, M., Gorkhover, T., Hirsemann, H., Potdevin, G., Graafsma, H., Nilsson, B., Chapman, H. N. & Hajdu, J. (2011). *Nature*. **470**, 78–81.

Shi, Y., Yin, K., Tai, X., DeMirci, H., Hosseinizadeh, A., Hogue, B. G., Li, H., Ourmazd, A., Schwander, P., Vartanyants, I. A., Yoon, C. H., Aquila, A. & Liu, H. (2019). *IUCrJ*. **6**, 331–340.

Sobolev, E., Zolotarev, S., Giewekemeyer, K., Bielecki, J., Okamoto, K., Reddy, H. K. N., Andreasson, J., Ayyer, K., Barak, I., Bari, S., Barty, A., Bean, R., Bobkov, S., Chapman, H. N., Chojnowski, G., Daurer, B. J., Dörner, K., Ekeberg, T., Flückiger, L., Galzitskaya, O., Gelisio, L., Hauf, S., Hogue, B. G., Horke, D. A., Hosseinizadeh, A., Ilyin, V., Jung, C., Kim, C., Kim, Y., Kirian, R. A., Kirkwood, H., Kulyk, O., Küpper, J., Letrun, R., Loh, N. D., Lorenzen, K., Messerschmidt, M., Mühlig, K., Ourmazd, A., Raab, N., Rode, A. V., Rose, M., Round, A., Sato, T., Schubert, R., Schwander, P., Sellberg, J. A., Sikorski, M., Silenzi, A., Song, C., Spence, J. C. H., Stern, S., Sztuk-Dambietz, J., Teslyuk, A., Timneanu, N., Trebbin, M., Uetrecht, C., Weinhausen, B., Williams, G. J., Xavier, P. L., Xu, C., Vartanyants, I. A., Lamzin, V. S., Mancuso, A. & Maia, F. R. N. C. (2020). *Commun. Phys.* **3**, 97.

Strüder, L., Epp, S., Rolles, D., Hartmann, R., Holl, P., Lutz, G., Soltau, H., Eckart, R., Reich, C., Heinzinger, K., Thamm, C., Rudenko, A., Krasniqi, F., Kühnel, K.-U., Bauer, C., Schröter, C.-D., Moshammer, R., Techert, S., Miessner, D., Porro, M., Hälker, O., Meidinger, N., Kimmel, N., Andritschke, R., Schopper, F., Weidenspointner, G., Ziegler, A., Pietschner, D., Herrmann, S., Pietsch, U., Walenta, A., Leitenberger, W., Bostedt, C., Möller, T., Rupp, D., Adolph, M., Graafsma, H., Hirsemann, H., Gärtner, K., Richter, R., Foucar, L., Shoeman, R. L., Schlichting, I. & Ullrich, J. (2010). *Nucl. Instr. Meth. Phys. Res. Sect. A* **614**, 483–496.

Thibault, P. & Menzel, A. (2013). *Nature*. **494**, 68–71.

Vartanyants, I. A. & Singer, A. (2010). *New J. Phys.* **12**, 035004.

Weierstall, U., Spence, J. C. H. & Doak, R. B. (2012). *Rev. Sci. Instrum.* **83**, 035108.

**Table 1**  Data sets selected at different stages of analysis: PSD fitting quality filtering, particle size filtering, single hit diffraction pattern selection. The percentage of the chosen data set to the initial one $S_0$ is given in parentheses.

| Analysis step | Data set name | Number of diffraction patterns |
|---|---|---|
| Initial data set | $S_0$ | $1.2 \times 10^7$ |
| Hit finding classification | $S_{hit}$ | 191,183 (1.6%) |
| PSD fitting score filtering | $S_{fit}$ | 179,886 (1.5%) |
| Particle size filtering | $S$ | 18,213 (0.1%) |
| 1st EM-based classification | $S_1^{EM}$ | 1,609 |
| 2nd EM-based classification | $S_2^{EM}$ | 1,402 |
| 3rd EM-based classification | $S_3^{EM}$ | 1,366 |
| 4th EM-based classification | $S_4^{EM}$ | 1,401 |
| 5th EM-based classification | $S_5^{EM}$ | 2,119 |
| Final EM-based classification | $S_{EM}$ | 1,085 (0.009%) |
| Manual selection | $S_{man}$ | 1,393 (0.01%) |

**Table 2**  Particle size analysis from facet to facet and from vertex to vertex for the AMO 2018 experimental data. All distances between facets and vertices are given in Supplementary Materials.

| | | Sizes |
|---|---|---|
| Facet-to-Facet | Mean | 61±2 nm |
| | Max | 64±2 nm |
| | Min | 59±2 nm |
| Vertex-to-Vertex | Mean | 63±2 nm |
| | Max | 67±2 nm |
| | Min | 60±2 nm |

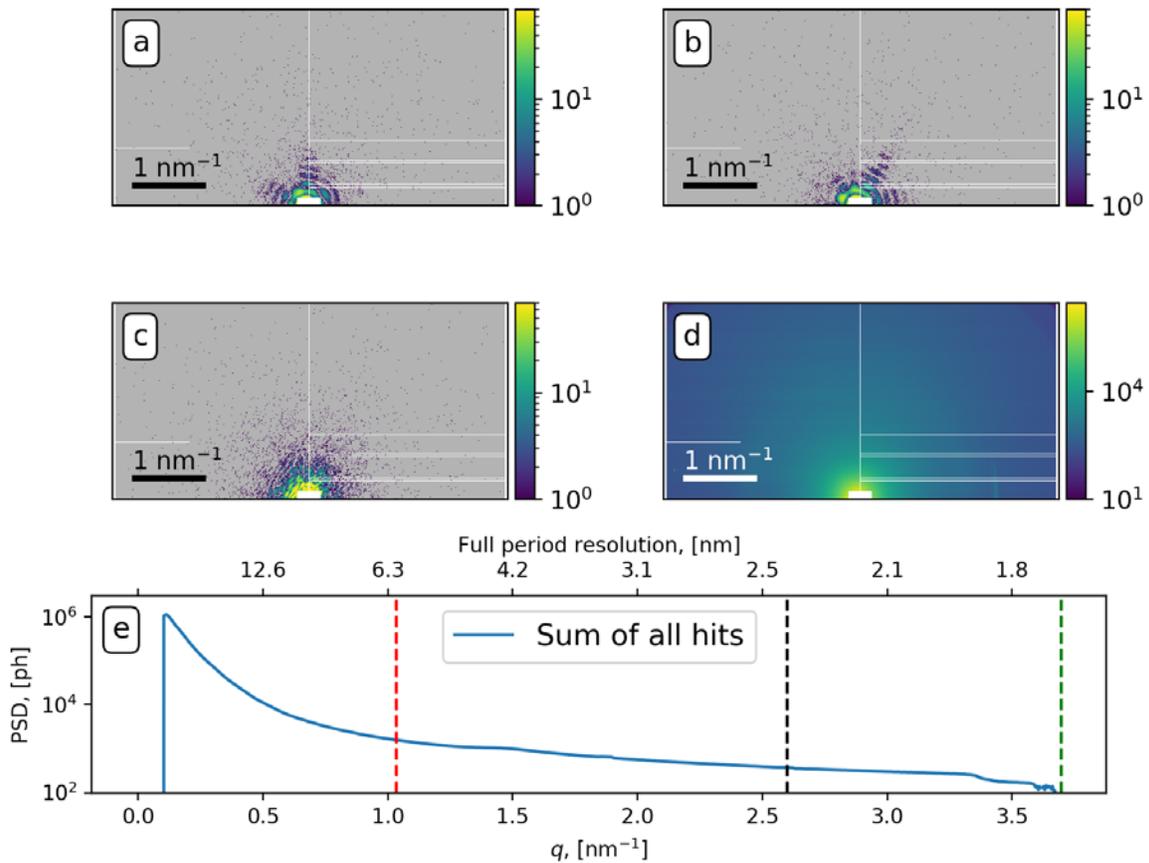

**Figure 1** Examples of diffraction patterns from the SPI experiment. (a) - (b) Diffraction patterns corresponding to a single PR772 virus hit by an XFEL beam. (c) Diffraction pattern corresponding to a non-single hit which was sorted out from the analysis at the classification step. (d) Sum of $1.9 \times 10^5$ diffraction patterns identified as hits. White regions in the center of diffraction patterns as well as white stripes correspond to a mask introduced to reduce artifacts due to the signal exceeding the detector capabilities. The mask was the same before and after the move of the detector panel. (e) PSD function of the scattered intensity for the sum of all diffraction patterns identified as hits collected in the experiment. The signal till the corner and the edge of the detector is indicated by the green and black vertical dashed lines, respectively. For orientation determination the data were used till momentum transfer values of one inverse nanometer (shown by red dashed line).

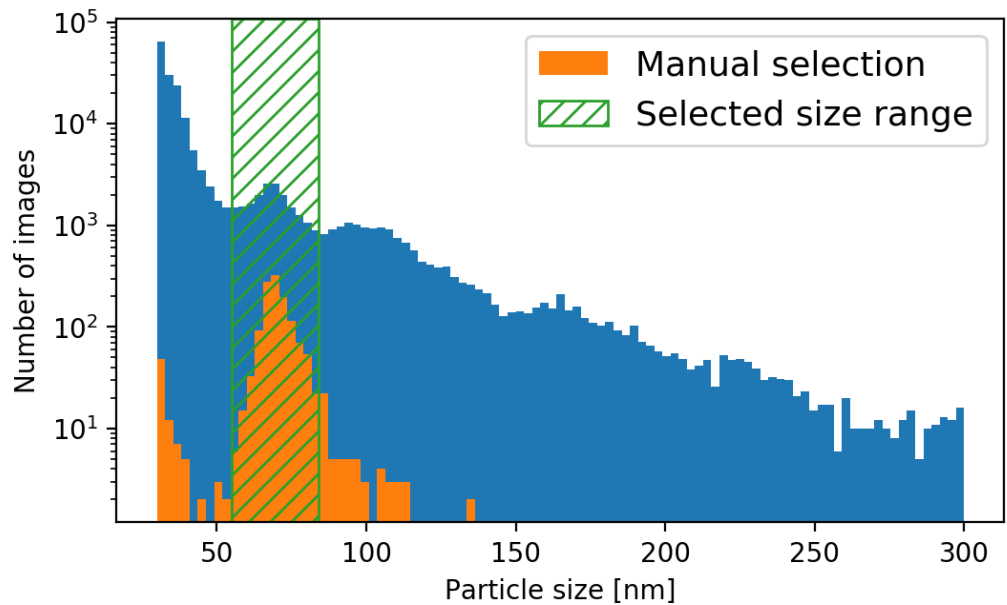

**Figure 2** Particle size histogram after the PSD fitting filtering. Blue area corresponds to $1.8\times10^{5}$ diffraction patterns satisfying the fidelity score criterion FS>1.05 (see Supporting Information for details). Orange area corresponds to manually selected single-hit diffraction patterns. Range of particle sizes from 55 nm to 84 nm ($1.8\times10^{4}$ diffraction patterns) was selected for the further SPI analysis (green dashed area).

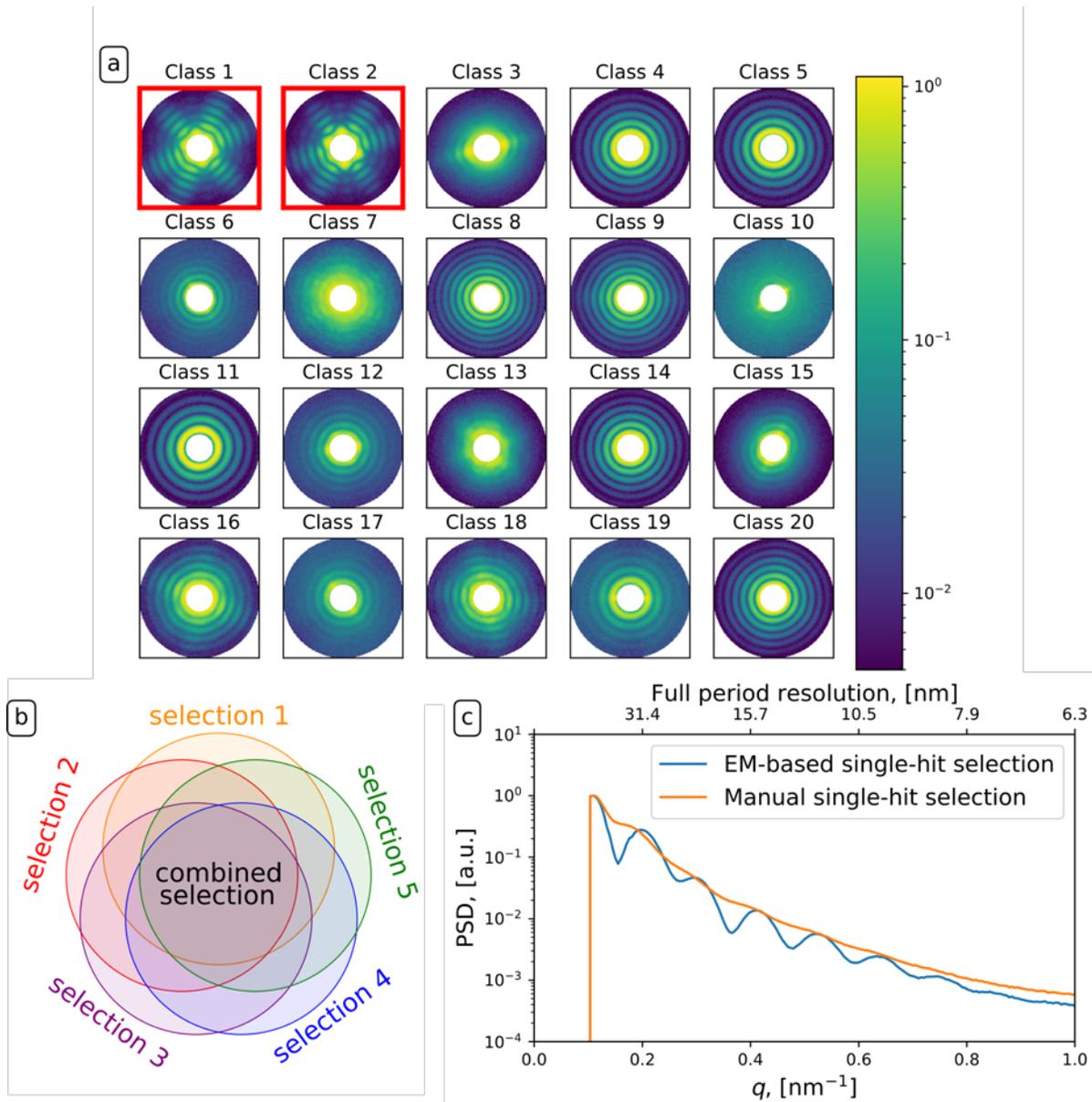

**Figure 3** Classification of diffraction patterns by expectation-maximization (EM) clustering. (a) Diffraction patterns are distributed into 20 classes according to their features. Classes 1 and 2 were selected as they clearly contain structural features of the investigated virus and its icosahedral shape. These two classes contain 1,609 diffraction patterns. (b) EM-clustering was repeated 5 times and intersecting selection with 1,085 patterns was considered for further analysis. (c) Averaged PSD functions for EM-based single-hit selection containing 1,085 patterns (blue line) and for manual selection containing 1,393 patterns (orange line).

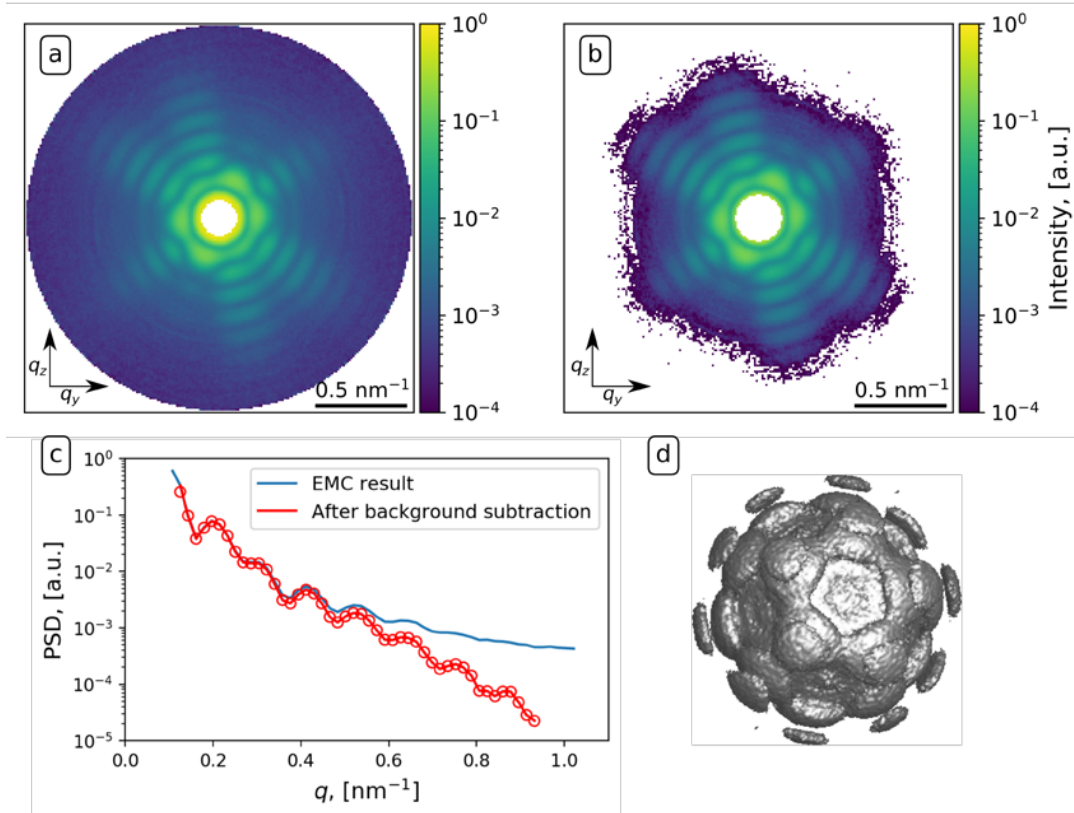

**Figure 4** Orientation determination results. (a) 2D $q_y$-$q_z$ cut of the 3D intensity distribution in reciprocal space, (b) the same intensity distribution after the background subtraction. (c) PSD function for the EMC-result (blue line) and after the background subtraction (red line with dots). It is well visible that the background subtraction enhanced structural visibility in the high $q$ region. (d) 3D intensity distribution in reciprocal space after the background subtraction shown at 0.5% level of the maximum value.

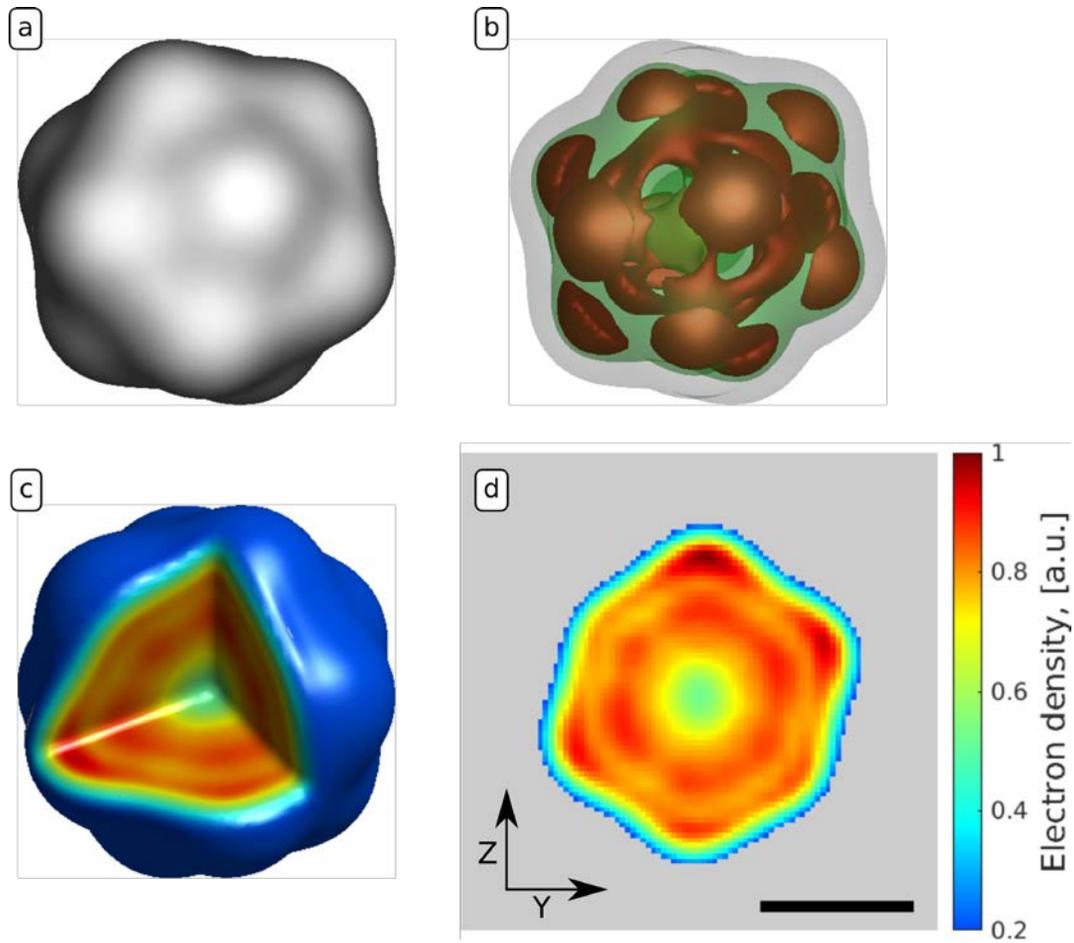

**Figure 5** Electron density of the reconstructed PR772 virus normalized to the maximum value. (a) Outer structure of the PR772 virus at the isosurface value of 20% of the maximum electron density. (b) Inner 3D structure of the PR772 virus at the isosurface values of 85% (brown area), 75% (green area), 20% (grey area). (c) 3D section of the virus. (d) Electron density slice of the virus. Amplitude values less than 0.2 were set to grey color scale. The color map for (c) and (d) is the same. Images are up-sampled three times for better visibility. The scale bar in (d) is 30 nm.

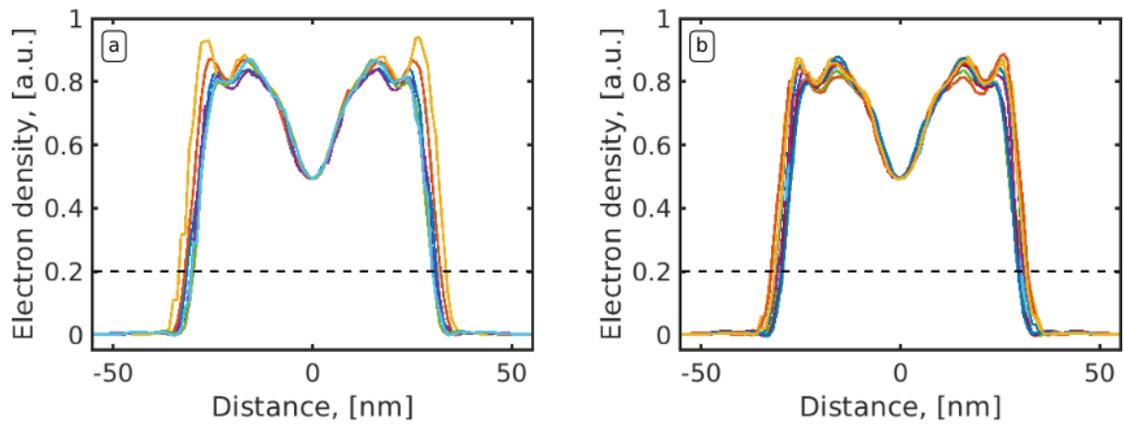

**Figure 6** Electron density profiles of the reconstructed virus PR772 normalized to the maximum value for the cut between vertices (a) and facets (b). Horizontal black dash line denotes particle size threshold of 0.2. The mean virus size is 63 nm and 61 nm for the distance between vertices and facets, respectively.

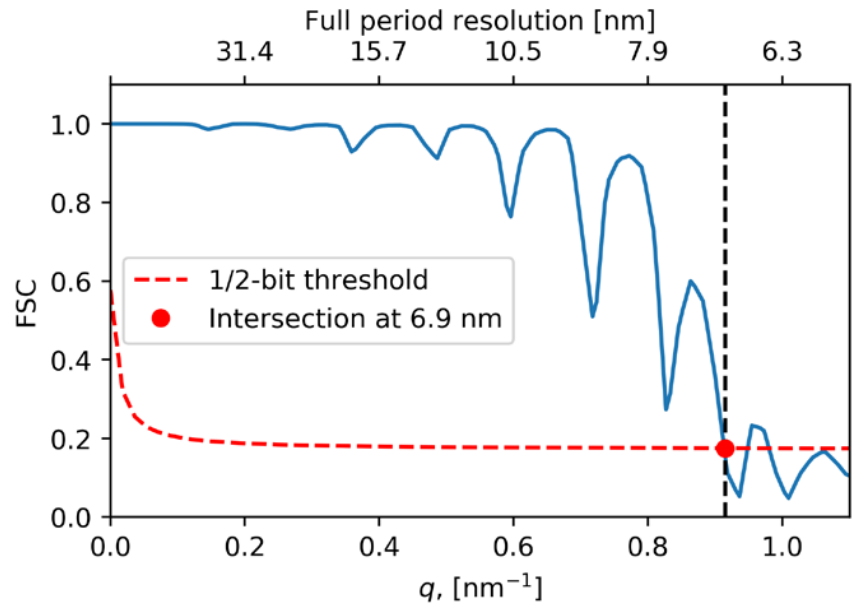

**Figure 7** Fourier shell correlation of the final reconstruction (blue line) that shows 6.9 nm resolution (red dot) with ½-bit threshold (red dashed line).

# Supporting information

### S1. Analysis of additional instrumental scattering

Visual inspection of the measured diffraction patterns showed an additional scattering signal close to the central part of the beam. This signal remains stable from pulse to pulse, which indicates that it most probably originates from beamline scattering. This additional instrumental scattering can be well seen on the averaged diffraction pattern in one of the experimental runs (see Fig. S1(a)).

We analyzed histograms of intensity for individual pixels and noticed that pixels with additional instrumental scattering most often recorded a signal of several photons. Contrary to that, pixels without this additional scattering most frequently recorded a signal of zero photons. We assumed that beamline scattering follows a Gaussian distribution and it was incoherently added to particle scattering. To correct this additional signal, we fit the first peak on histogram of intensity for each pixel by a Gaussian function (see Fig. S2). Then we subtract the value of the Gaussian center from the total signal of this pixel for all diffraction patterns. This instrumental scattering subtraction was crucial for further beam center position finding and particle size filtering. We did not mask this region because we would lose important information about the first diffraction minimum.

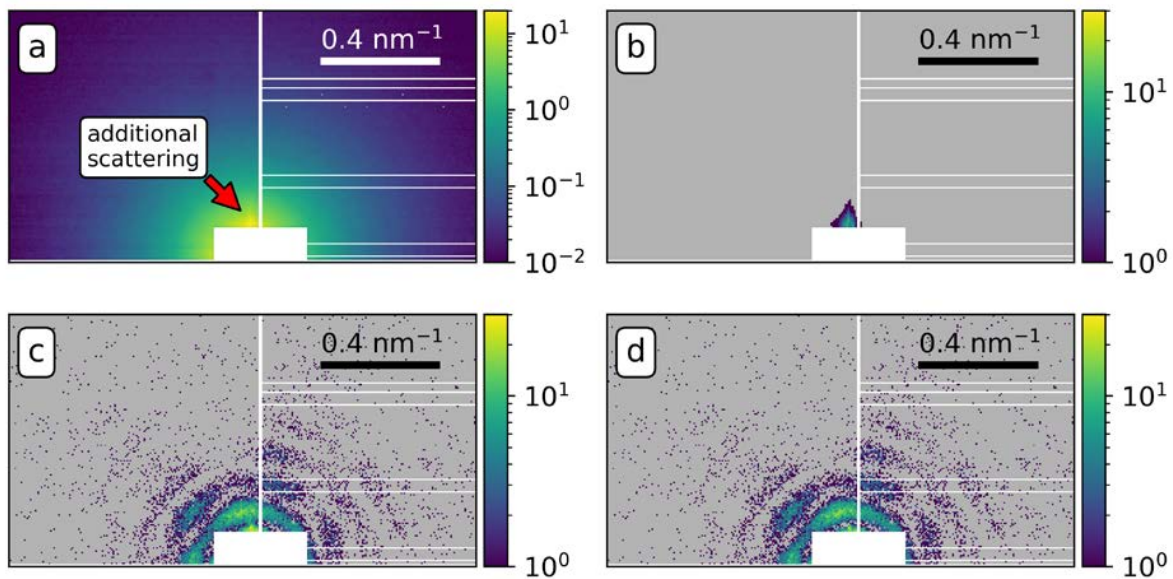

**Figure S1** (a) An averaged diffraction pattern of one of the runs. White regions in the diffraction patterns correspond to a mask introduced to hide misbehaving pixels. Additional instrumental scattering originating from the beamline is well visible in the central part of the averaged diffraction pattern. (b) Identified additional scattering for this run. Diffraction pattern before (c) and after (d) subtraction of additional scattering.

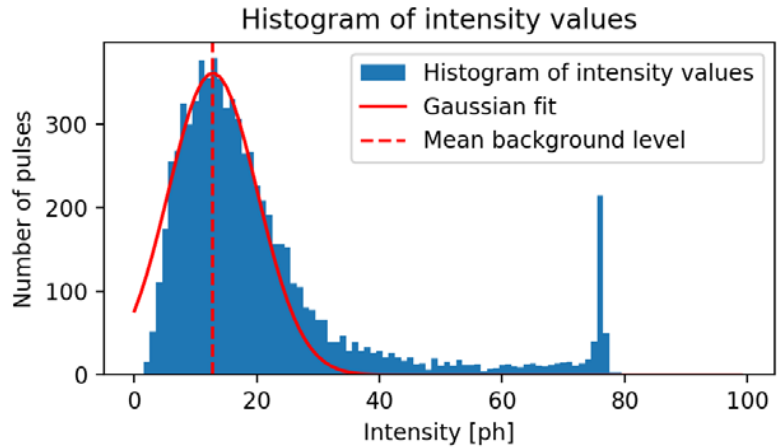

**Figure S2** Histogram of intensity values for a selected pixel from one of the runs with strong additional instrumental scattering. The pixel shows the most frequently recorded value is 13 photons. This histogram was fitted with a Gaussian function and the mean value of this Gaussian function was subtracted from all intensity values for different pulses corresponding to that pixel. The peak on the right side of the histogram is due to limitations of the detector: if the detector pixel collects more than 75 photons, its response is always in the range of 76-79 photons.

## S2. Beam center position finding

In the present work, the beam center position was retrieved from the diffraction patterns. Detector consists of two identical panels with the gap between them for direct beam propagation. Due to the fact that signal from only one panel was available, the beam center position could not be determined by centrosymmetric property of diffraction patterns. Furthermore, the detector panel was moved during the experiment, and we estimated the beam position twice – before and after the detector panel was moved. The beam center position was determined in the following way. First, the sum of all diffraction patterns at an arbitrary chosen fixed position of the detector was calculated. The resulting average diffraction pattern was rotationally symmetric and allowed a rough estimate of the beam position center. For the success of this step, it was crucial to subtract parasitic scattering from the beamline as described in the previous section. To define the beam position center more carefully on the next step, diffraction patterns with a narrow distribution of particle sizes were selected and the averaged diffraction pattern was obtained. This diffraction pattern has pronounced diffraction fringes and it was correlated with the two-dimensional (2D) form factor of a spherical particle (see Fig. S3). Inspection of this method on simulated data with similar parameters showed that mean deviation of the refined center from the true center of diffraction patterns is less than half of a pixel.

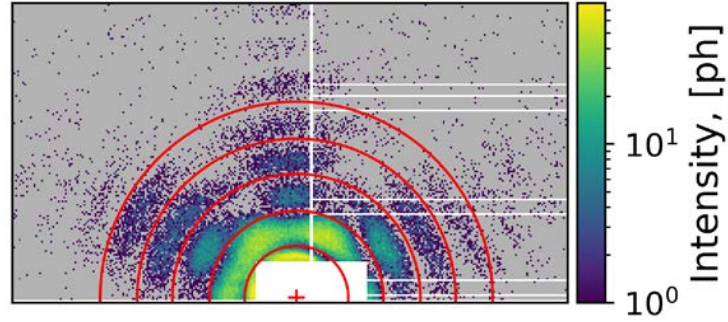

**Figure S3** Center position on a selected diffraction pattern. Minima of the optimal spherical form factor are shown by red circles.

### S3. Particle size filtering

The particle size filtering was based on the fitting of the power spectral density (PSD) function of each diffraction pattern with the form factor of a sphere. A set of form factors corresponding to the spheres with the diameter in the range from 30 to 300 nm was generated first. On the next step the PSD function of each diffraction pattern was fitted with a spherical form factor function from the generated set (see Fig. S4(a)). As the fit quality measure for the certain size (diameter) of the spherical particle, the mean difference was used

$$D_s = \frac{1}{q_{max} - q_{min}} \sum_{q_{min}}^{q_{max}} \left| I_{exp}(q) - I_s(q) \right|, \quad (S1)$$

where $I_{exp}(q)$ is the PSD value of the experimental intensity for selected $q$, $I_s(q)$ is the form factor of a sphere with the size (diameter) S. In equation (S1) the $q$-values were ranging from $q_{min}$=0.12/0.15 nm$^{-1}$ before and after the detector panel was moved, up to $q_{max}$=0.66 nm$^{-1}$. An example of the mean difference function of equation (S1) obtained for one of the diffraction patterns is shown in the Fig. S4(b). This function has several minima, where the first minimum corresponds to a sphere with the best size. The second minimum corresponds to a sphere with the second-best size, *etc.* To measure fidelity of the particle size estimation we used fidelity score (FS) defined as

$$FS = \frac{D_{S_2}}{D_{S_1}}, \quad (S2)$$

where $D_{S_1}$ and $D_{S_2}$ are the values of the mean difference function $D_S$ corresponding to the first and second minima in Fig. S4(b). The fidelity score histogram for all diffraction patterns identified as hits ($1.9 \times 10^5$ diffraction patterns) is shown in Fig. S5. According to its definition in Eq. (S2), if the fidelity score is equal to unity ($FS = 1$) it means that $D_S$ values equal for two different minima, therefore fitting cannot find an appropriate size for a particle that will correspond to a given diffraction pattern. We introduced a threshold value of $FS = 1.05$ and considered all diffraction patterns with the fidelity score higher than this value (see Fig. S5). By that we determined $1.8 \times 10^5$ diffraction patterns that were selected for the further particle size filtering described in the main text.

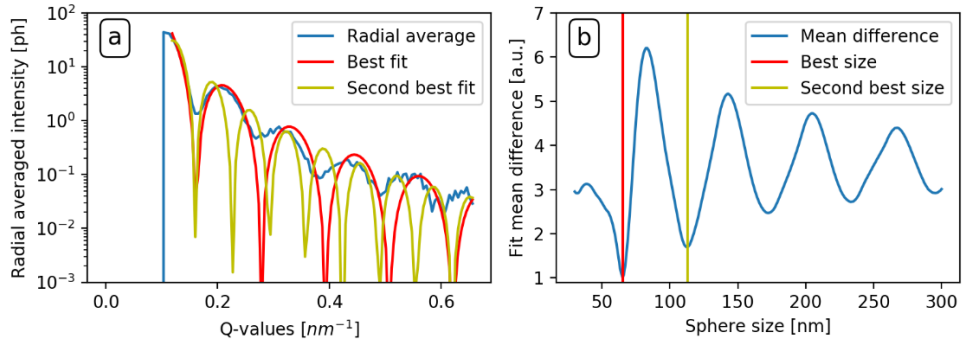

**Figure S4** PSD fitting analysis of the diffraction pattern. (a) PSD function (blue line) was fitted with the form factors of spherical particles of different size. Red and yellow lines correspond to the form factors of the spherical particles with the best and second best size, that were used for calculation of fidelity score. (b) Mean difference function as defined in Equation (S1). Fidelity score value is 1.6 for this diffraction pattern.

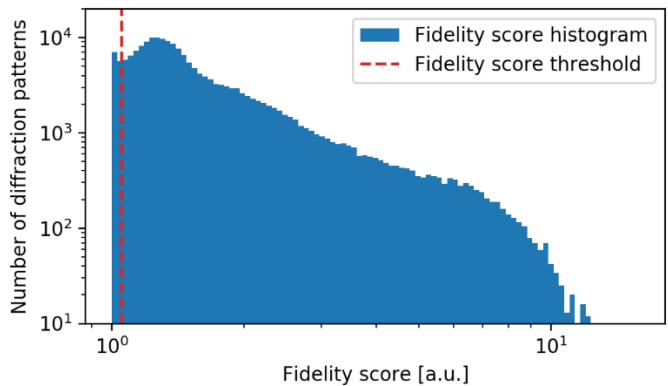

**Figure S5** Fidelity score histogram for all diffraction patterns identified as hits in the experiment. Fidelity score threshold of the value 1.05 is shown as the vertical dashed red line. $1.8 \times 10^5$ selected diffraction patterns with fidelity score above threshold were used for further analysis.

Virus size distribution according to the cryo-EM studies of the PR772 virus used in the experiment is shown in Fig. S6.

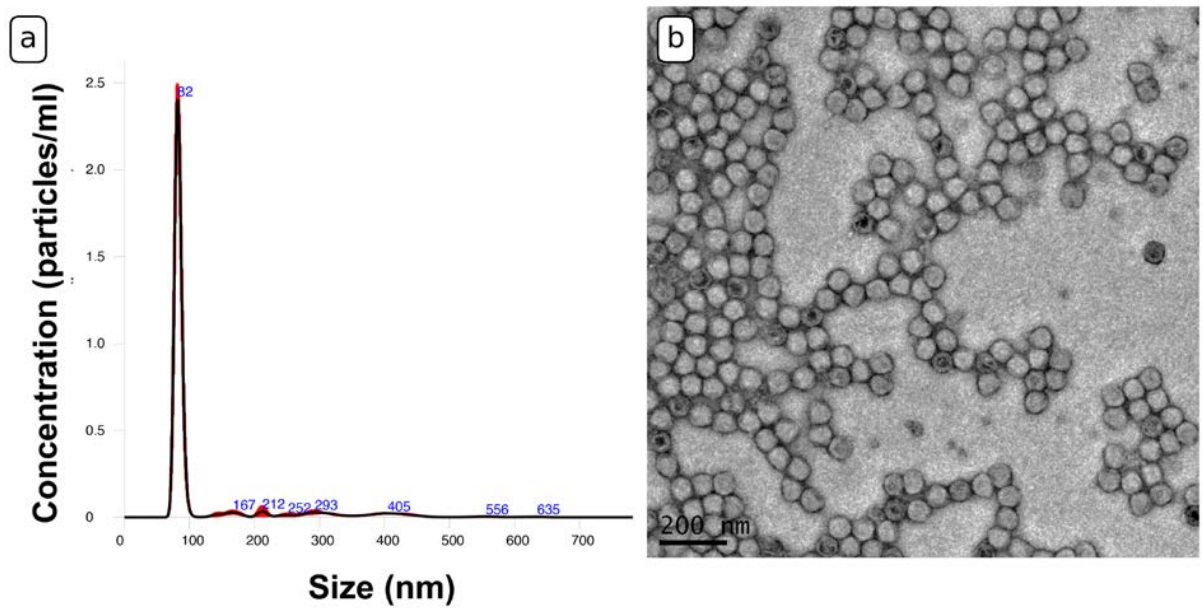

**Figure S6** CryoEM structural studies of PR772 used in the experiment. (a) Virus size distribution profile. (b) PR772 visualization with screening transmission electron microscopy (TEM).

After the size filtering and running EM algorithm, we ended with the data set containing 1,085 patterns (see Table 1 of the main text). In order to identify performance of single particle collection as well as efficiency of the 3D printed nozzles we plot a histogram of selected patterns as a function of experimental run (see Fig. S7). This histogram shows that collection was significantly improved towards the end of the experiment.

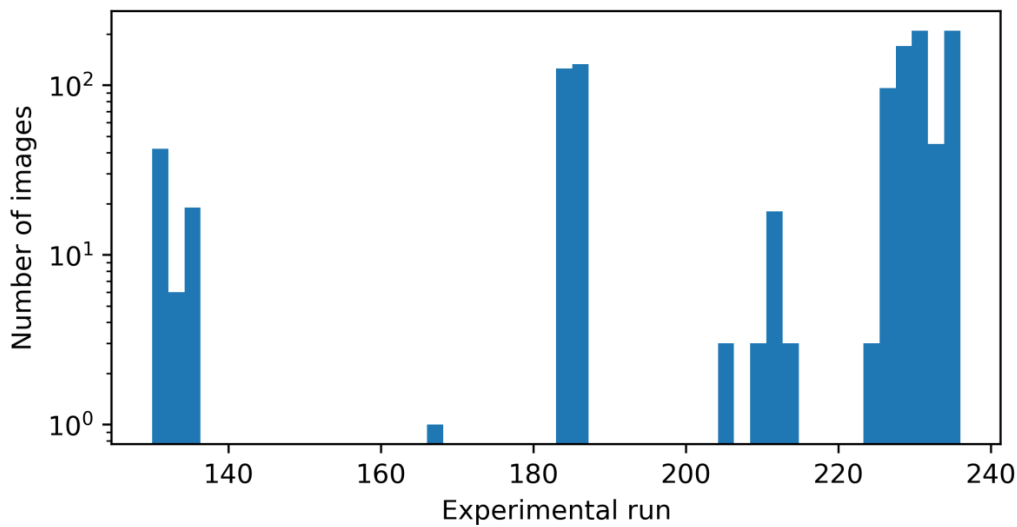

**Figure S7** Histogram of the number of images per experimental run from the data set containing 1,085 patterns.

## S4. Orientation determination and background subtraction

Before orientation determination we defined the region of interest for the collected data. This was performed twice, before and after the detector panel was moved. For the first/second detector position the detector area from $q_{min}$=0.18/0.23 nm$^{-1}$ to $q_{max}$=1.03 nm$^{-1}$ was considered. The data at $q < q_{min}$ was excluded from orientation determination due to poor convergence, but these data were used to compute the 3D amplitude in reciprocal space. The data corresponding to $q > q_{max}$ were removed from the analysis because the scattering signal is indistinguishable from noise. The selected data was binned 2 by 2 due to computing memory constraints. The diffraction patterns were then converted into Dragonfly (Ayyer *et al.*, 2016) input format using exact coordinates of each pixel, two retrieved beam center positions (before and after detector panel was moved), and other experimental parameters *such as* the wavelength of 7.3 Å and the distance from the interaction region to the detector of 130 mm. Orientation determination was performed with quaternion sampling of 10, the starting beta value 0.01 which was multiplied by 1.3 every 30 iterations. We performed 300 iterations with the EMC algorithm and took iteration 270 as the best one (Fig. S8).

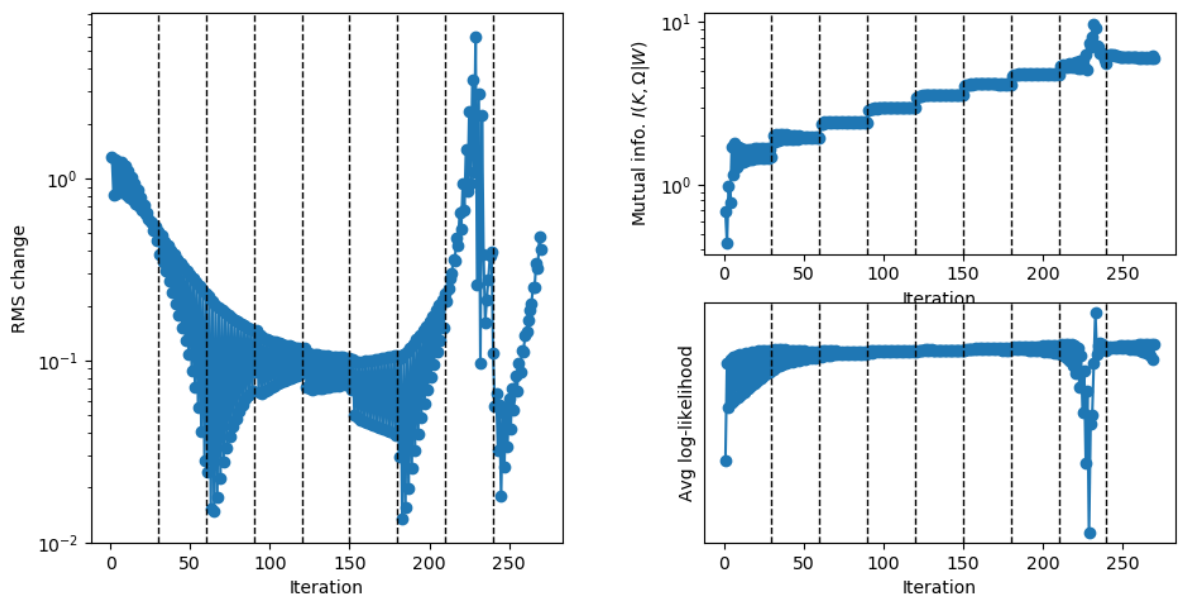

**Figure S8** Screenshot of 'Dragonfly' software showing diagnostics of each 3D orientation reconstruction. (a) R.M.S. change indicates the degree of model modification at each iteration. (b) Mutual information between model tomograms and experimental data as the metric of the 'sharpness' of the probability distribution over orientations. (c) Average log-likelihood of patterns given a model as a metric of how an iterative reconstruction approaches the global likelihood maximum.

Results of the EMC algorithm for orientation determination are shown in Fig. S8(a-c). It is well seen that at high $q$-signal level some background is still present after orientation determination.

To determine the background level, the signal in selected areas of the high $q$-region, where the contribution of the meaningful data is minimal, was analyzed (see Fig. S9(a-c)). The histogram of intensity in this area is shown in Fig. S9(d). The background level was defined as the mean signal value and was subtracted from the 3D intensity map in reciprocal space. Negative values of intensity in the final representation were set to zero.

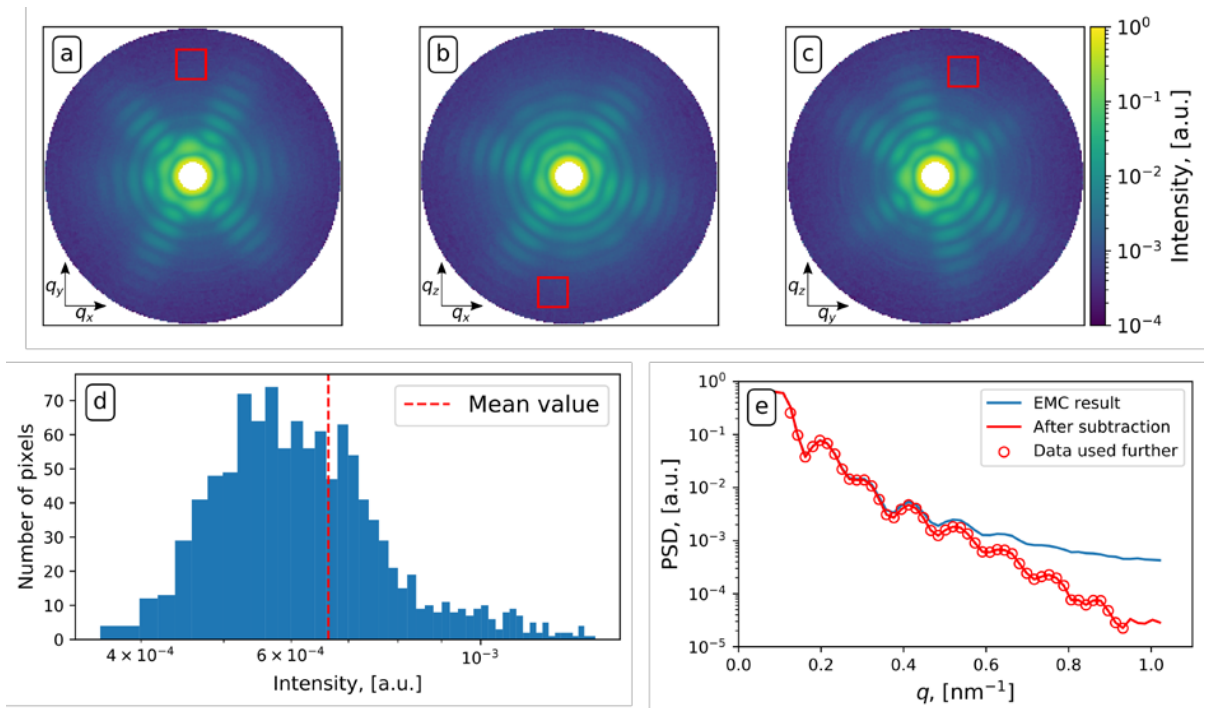

**Figure S9** Results of the EMC orientation determination algorithm. (a-c) Orthogonal two-dimensional cuts through the center of the 3D volume of reciprocal space after application of the EMC algorithm. For the background estimate the intensity values in the region of high $q$ shown with red squares were analyzed. (d) Histogram of the signal from the area shown in (a-c). The mean value of the signal is shown with the vertical dashed red line. (e) PSD functions before (blue line) and after (red line) background subtraction. To avoid artifacts at low and high $q$-values a part of the curve indicated with red dots was considered for further analysis.

The PSD function after background subtraction (red line in Fig. S9(e)) reveals artifacts in the regions of low ($q < 0.12$ nm$^{-1}$) and high ($q > 0.93$ nm$^{-1}$) momentum transfer values. Since the data in these regions did not follow the expected spherical form factor behavior, we did not consider this part of the data. Data used for further analysis are shown with red dots in Fig. S9(e). For visual inspection we show the final 3D intensity distribution of the PR772 virus in animation.

## S5. Phase retrieval

Phase retrieval and electron density determination of the PR772 virus was performed in several steps. First, due to masking, the central part of the 3D intensity map was missing (see Fig. S9(a-c)). To recover it, several reconstructions, with an assumption of free-evolving intensity in this part of reciprocal space, were performed. The following algorithms were considered at this stage: 90 iterations of cHIO with the feedback value 0.8 (Fienup, 2013), 200 iterations of the ER algorithm (Fienup, 1982) with alternation of the shrink-wrap algorithm each 10 iterations with the threshold value of 0.2 and Gaussian filtering with 3 to 2 sigma. (Marchesini *et al.*, 2003). This combination of algorithms was repeated three times for one reconstruction with the total number of 870 iterations. All obtained reconstructions showed identical central part and we used one of them in further analysis. In Fig. S10(a) the PSD functions of the initial and one of the reconstructed data are shown. One can see from that figure that the reconstructed curve follows very well the experimental data points. For the low $q$-values below 0.14 nm$^{-1}$ the experimental data were substituted with the data obtained in phase retrieval. Difference between experimental data and reconstruction in the central fringe is contributed to incorrectly reduced detector signal for intensity above 75 photons (Fig. S2). This modified 3D intensity map was used for the final phase retrieval and virus structure determination (Fig. S10(b)).

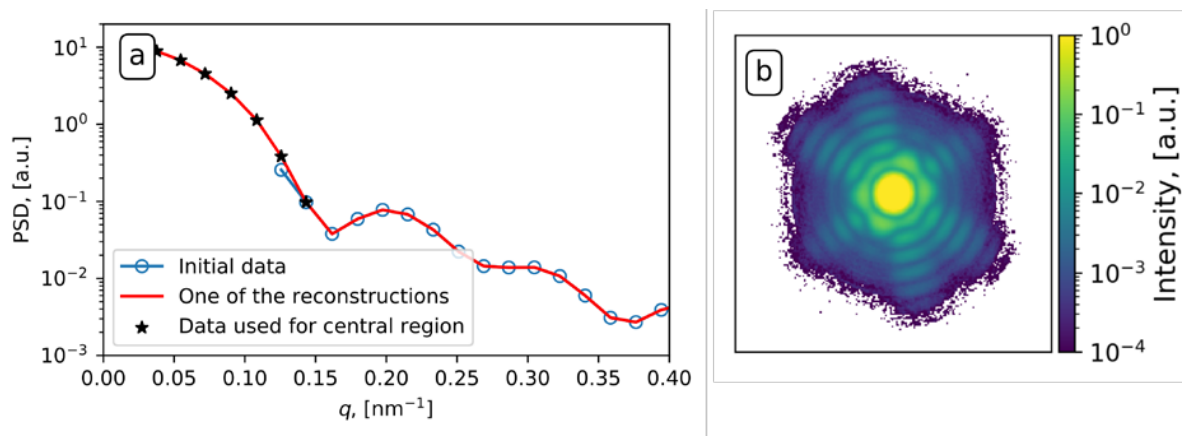

**Figure S10**(a) PSD functions for experimental data (blue empty dots) and one of the reconstructions (red line). The central part below $q = 0.14$ nm$^{-1}$ (black stars) was taken from this reconstruction for further analysis (b) Modified data with the filled central part. White area around the diffraction pattern is the part of reciprocal space where the data were set to zero initially but were allowed to freely evolve during the iterative phase retrieval.

On the next step 50 individual reconstructions were performed. In these reconstructions intensities at high $q$-values (in the regions where they were initially set to zero (white area in Fig. S10(b)) were allowed to freely evolve with a weight factor of 0.9. The initial support was taken as a Fourier transform of the 3D data used for reconstructions and had spherical shape with diameter about 90 nm. The same sequence of algorithms was used for these reconstructions as mentioned above plus it was performed 100 iterations of the Richardson-Lucy algorithm (Clark *et al.*, 2012) with the total number of 970 iterations. This algorithm based on deconvolution technique allowed to additionally enhance the contrast of the reconstructed diffraction patterns to the value of ⟨γ⟩=0.87, and by that remove the remaining background from the 3D diffraction patterns, which is defined as point spread function (PSF) shown in Fig. S11. As a result, we obtained complex valued real space images for each 50 reconstructions from which the absolute value was considered as an electron density of the virus.

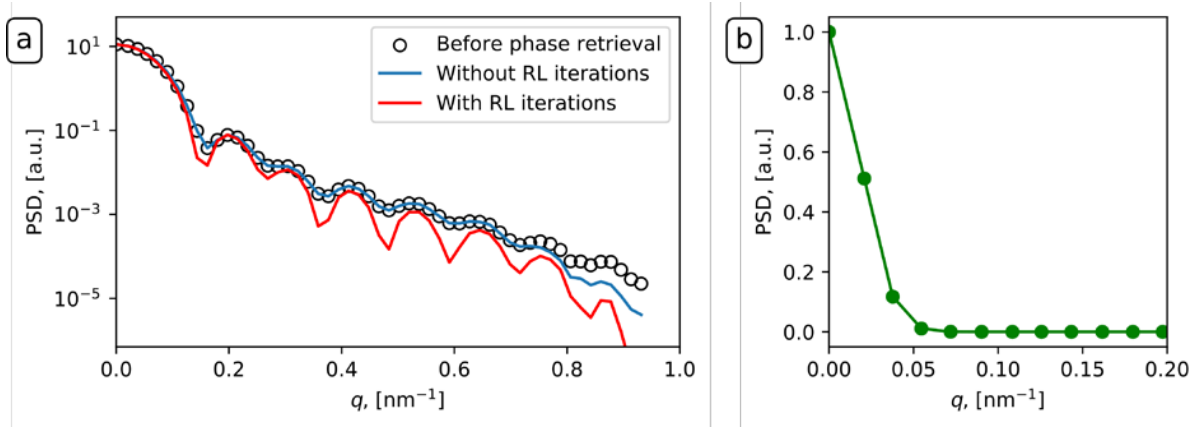

**Figure S11** (a) PSD functions for data before phase retrieval (black empty dots) and one of the reconstructions without Richardson-Lucy iterations (blue line) and with them (red line). This additional deconvolution allowed us to improve contrast of diffraction patterns to the value of ⟨γ⟩=0.87. (b) PSF for individual reconstruction as a result of Richardson-Lucy deconvolution algorithm. Intensity is normalized to the maximum value.

### S6. Mode decomposition and virus electron density analysis

To determine the final electron density of the virus, the mode decomposition of the reconstruction set was performed and as an outcome of this procedure an orthogonal set of modes was found. The whole procedure consists of the following steps (see Fig. S12 and (Khubbutdinov *et al.*, 2019)):

a) Initial 4D matrix (Fig. S12(a)) consists of 3D amplitudes of the reconstructions (203x203x203 pixels), where the fourth dimension is given by the number of reconstructions (50 in the present case).

b) This 4D matrix of amplitudes is rearranged into a 2D matrix (Fig. S12(b)) with 50 columns, where each 3D amplitude matrix was rearranged to a 1D column.

c) Next, the mode decomposition is performed for the density matrix that is obtained by multiplication of the previously defined 2D matrix transposed complex conjugated and 2D matrix itself (Fig. S12(c)). By diagonalization of this matrix using Principal Component Analysis (PCA), eigenfunctions and eigenvalues of the reconstructed object are obtained.

We considered the fundamental mode of the reconstruction set (with a weight of 99%) as a final result.

The final electron density of the PR772 virus was three times up-sampled for better visual impression. The comparison between the initial and up-sampled structures is shown in Fig. S13. The electron density of the reconstructed PR772 virus was normalized to the maximum value in this figure. For visual inspection we also show the final virus structure (outer, inner and through x-axis) in animations.

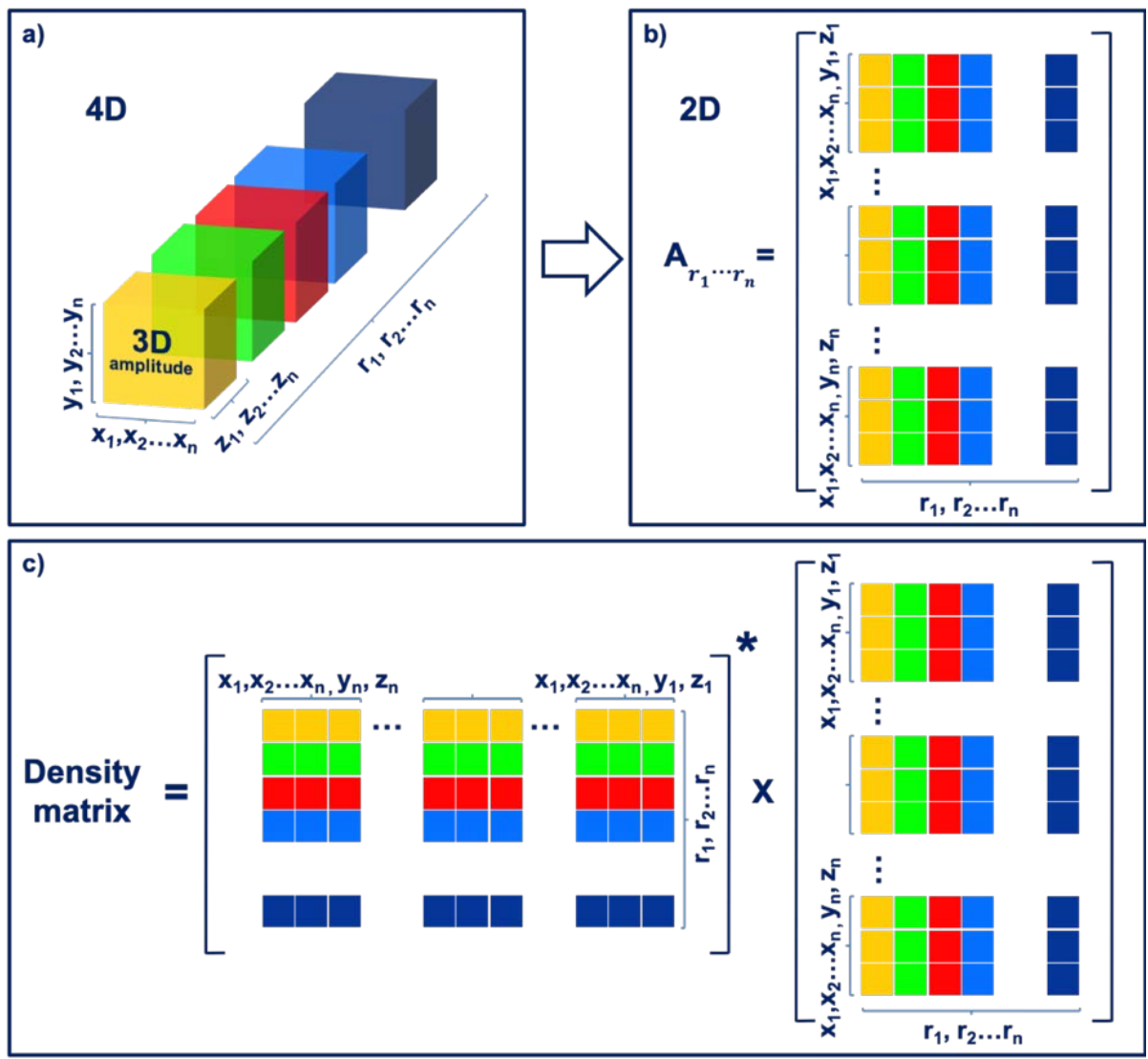

**Figure S12** Mode decomposition procedure for the set of the reconstructions obtained by phase retrieval. (a) Initial 4D matrix consisting of 3D amplitudes of the reconstructions (203×203×203 pixels), where the fourth dimension is the number of reconstructions. (b) 4D matrix rearranged to 2D matrix, where each 3D amplitude matrix was rearranged to 1D column, the number of columns corresponds to the number of reconstructions. (c) Density matrix obtained by the multiplication of 2D matrices (b): its transposed complex conjugated and matrix itself.

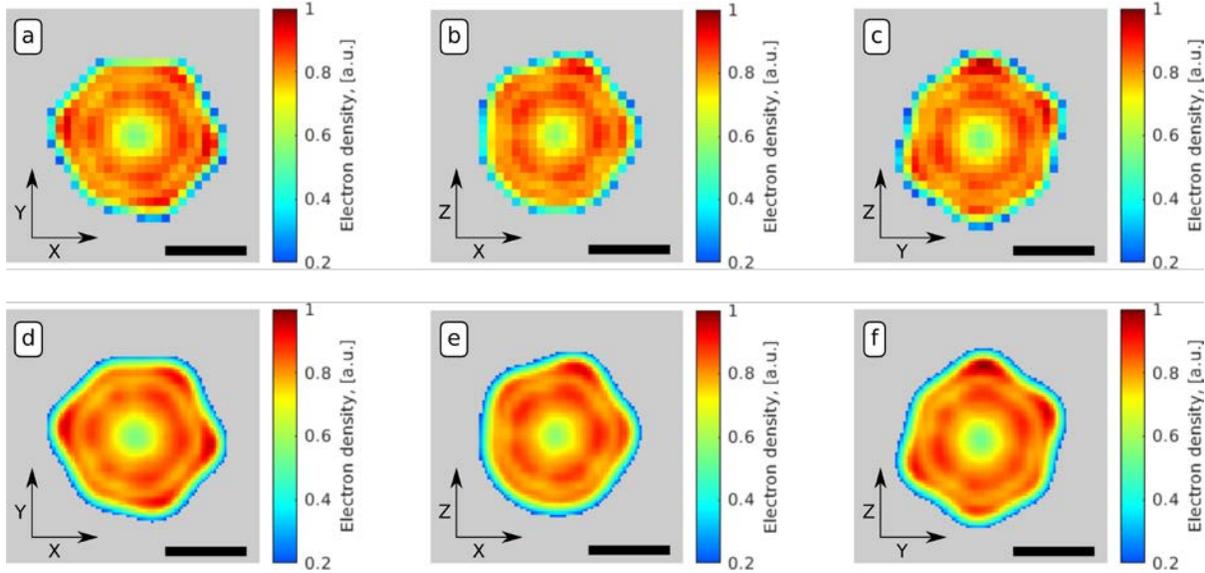

**Figure S13** Final electron density of the virus obtained as a result of mode decomposition. Black line denotes 30 nm. (a-c) Results of the initial reconstruction. (d-f) Three times up-sampled results from (a-c). Electron density less than 0.2 was set to grey color scale.

The virus size was obtained from the analysis of the electron density profiles. For the particle size estimate we selected the electron density threshold value of 0.2 as it was considered in the shrink-wrap algorithm during the phase retrieval. From this criterion we determined the particle sizes in the directions from facet to facet and from vertex to vertex (Rose *et al.*, 2018), which are shown in Table S1. The mean particle size was $61 \pm 2$ nm (between facets) and $63 \pm 2$ nm (between vertexes).

**Table S1**   The virus sizes in the directions from facet to facet and from vertex to vertex. The mean sizes in each direction are shown in the last column.

| | | | | | | | | | | | Mean size |
|---|---|---|---|---|---|---|---|---|---|---|---|
| Facets, nm | 62±2 | 60±2 | 64±2 | 62±2 | 60±2 | 60±2 | 60±2 | 59±2 | 64±2 | 63±2 | 61±2 |
| Vertexes, nm | 63±2 | 64±2 | 67±2 | 61±2 | 60±2 | 61±2 | | | | | 63±2 |

To estimate the capsid size, we analyzed a few electron density profiles (see main text Fig. 6(a-b) and Fig. S14), which were fitted with four Gaussian functions. The area of fitting was considered according to the electron density threshold 0.2, similar to the shrink-wrap value during the reconstruction. Fitting the result for the electron density profile is shown in Fig. S14. Left and right Gaussian functions correspond to the capsid part of the virus structure. Taking the FWHM values of these curves we determined the capsid size to be $7.6 \pm 0.3$ nm.

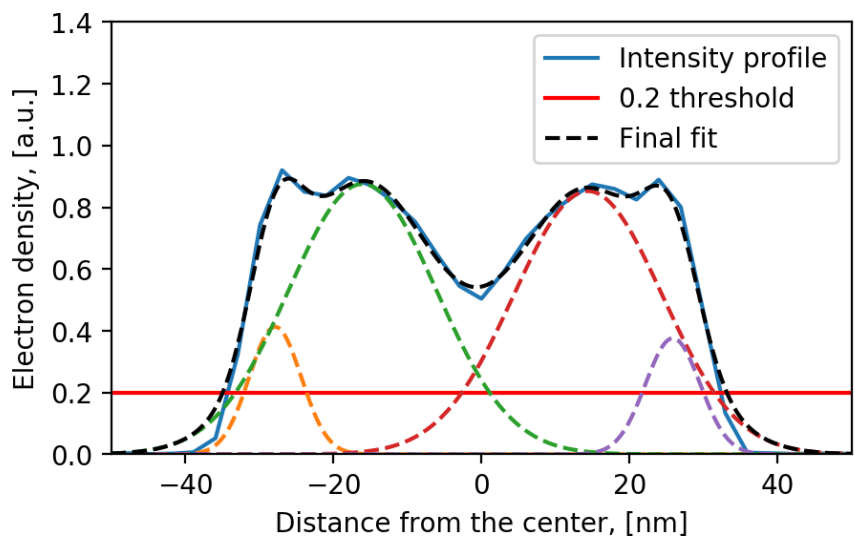

**Figure S14** Analysis of the electron density profile. For the capsid size estimate, к fitting of the electron density line profile with four Gaussian functions was performed. Left and right (orange and purple) Gaussian functions correspond to the capsid. The mean size of the capsid was determined as FWHM of these Gaussian functions and is equal to $7.6 \pm 0.3$ nm.

## References

Ayyer, K., Lan, T.-Y., Elser, V. & Loh, N. D. (2016). *J. Appl. Crystallogr.* **49**, 1320–1335.

Clark, J. N., Huang, X., Harder, R. & Robinson, I. K. (2012). *Nat. Commun.* **3**, 1–6.

Fienup, J. R. (1982). *Appl. Opt.* **21**, 2758–2769.

Fienup, J. R. (2013). *Appl. Opt.* **52**, 45–56.

Khubbutdinov, R., Menushenkov, A. P. & Vartanyants, I. A. (2019). *J. Synchrotron Radiat.* **26**, 1851–1862.

Marchesini, S., He, H., Chapman, H. N., Hau-Riege, S. P., Noy, A., Howells, M. R., Weierstall, U. & Spence, J. C. H. (2003). *Phys. Rev. B.* **68**, 140101.

Rose, M., Bobkov, S., Ayyer, K., Kurta, R. P., Dzhigaev, D., Kim, Y. Y., Morgan, A. J., Yoon, C. H., Westphal, D., Bielecki, J., Sellberg, J. A., Williams, G., Maia, F. R. N. C., Yefanov, O. M., Ilyin, V., Mancuso, A. P., Chapman, H. N., Hogue, B. G., Aquila, A., Barty, A. & Vartanyants, I. A. (2018). *IUCrJ.* **5**, 727–736.